\documentstyle[12pt,aasms4]{article}
\lefthead{Kenji Bekki}
\righthead{Formation of a polar-ring galaxy}
\begin{document}
\title{Formation of a polar-ring galaxy in a galaxy merger} 
\author{Kenji Bekki}
\affil{Astronomical Institute, Tohoku University, Sendai, 980-77, Japan \\
email address: bekki@astroa.astr.tohoku.ac.jp}

\begin{abstract}
  We numerically investigate stellar and gas dynamics in star-forming
  and dissipative galaxy mergers between
 two disk galaxies  with specific orbital  configurations.
  We find that violent relaxation combined with gaseous dissipation   
 in galaxy merging transforms two disk galaxies into one S0 galaxy with polar-rings:
 Both the central S0-like host and the polar-ring component in a polar-ring galaxy
 are originally  disk galaxies. 
 We also find that morphology of the developed polar-rings reflects both
 the initial orbit configuration of galaxy merging and
 the initial mass ratio 
 of the two merger progenitor disk galaxies.
 Based upon these results, we discuss the origin of the
 fundamental observational 
 properties of
 polar-ring galaxies, such as the prevalence of S0 galaxies among polar-ring galaxies,
 the rarity of polar-ring galaxies among S0 galaxies, the dichotomy between narrow 
 polar-rings and annular ones, shapes of polar-ring warps,
 and an appreciably larger amount
 of interstellar gas in the polar-ring component.
\end{abstract}

\section{Introduction}
 Polar-ring galaxies are generally considered to be dynamically peculiar systems in which 
 the outer rings composed of gas and stars are aligned roughly in a perpendicular orientation  with
 respect to the major axis of the central host galaxies (Schweizer, Whitmore, \& Rubin 1983;
 Whitmore et al. 1990; Sackett 1991).
 A growing number of observational studies 
 have been recently accumulated which can provide valuable
 information about the origin of these peculiar polar-ring galaxies.
 Nearly all of the  central host  are morphologically
 normal S0 galaxies, some of which are confirmed to be rapidly rotating by kinematical 
  studies (Schechter \& Gunn 1978; Schechter, Ulrich, \& Boksenberg 1984;
  Whitmore et al. 1990; Whitmore 1991).
 Approximately only 0.5 percent of all S0 galaxies have observable polar-rings, which suggests
 that a particular mechanism is required for the formation 
 of polar-rings in S0  galaxies. 
 The  ring component also shows
 rapid rotation comparable to that of the main host galaxies, implying that two dynamically
 different system coexist in these polar-ring galaxies.
 An appreciable amount of HI gas, which is sometimes comparable to the 
 total mass of the host,  is closely associated 
 with the stellar ring  component (e.g., Shane 1980; Schechter et al. 1984;
 Richter, Sackett, \& Sparke 1994;
 Arnaboldi et al. 1997; Galletta, Sage, \& Sparke 1997).
 The morphology of the polar-rings is basically divided into two broad classes
 (Whitmore 1991): A narrow ring  which 
 is not extended in size  (e.g., ESO 415-G 26) and an annulus which is a disk-like component 
 with  the central part cut out (e.g., NGC 4650a).
 Peculiar morphology is  observed in some polar-ring galaxies (e.g., the Helix galaxy,
NGC 2685,  and double ringed system, ESO 474-G 26),
which further implies  considerably
complicated physical processes in  polar-ring formation 
and simultaneously provides a clue
to the understanding of the origin of polar-ring galaxies (Sackett 1991).   
 Roughly two-thirds of these polar-rings show  obvious galactic  warps
 whose shapes look like `integral
 sign' and/or   `banana'(Whitmore 1991). 
 Statistical studies on the distribution of the angle between the ring component and the central 
 host reveal that these two components strongly prefer to be orthogonal with each  other.
 These peculiarities both in the kinematics and morphology observed in
 polar-ring galaxies have attracted
 a number of  theoretical interests, which are divided basically into two categories:
One is the origin of the polar-rings and the other is the nature of dark matter halo
 surrounding  polar-ring  galaxies.
Although there are
a large number of important studies addressing the three dimensional shapes
of dark matter halo in the galaxies (e.g., Whitmore, Mcelroy, \& Schweizer 1987;
Reshetnikov \& Combes 1994; Sackett et al. 1994; Combes \& Arnaboldi 1996),  
 we here restrict ourselves to the mechanisms which would naturally explain 
 the formation of the polar-ring galaxies with spherical haloes.

 It is generally believed that the formation of the polar-rings is 
 the results of 
  a `secondary event' involving  
   a pre-existing S0 galaxy (e.g., Steiman-Cameron \& Durisen 1982; Sparke 1986;
 Quinn 1991; Rix \& Katz 1991; Reshetnikov \& Sotnikova  1997).
 Specifically, the host S0 galaxy is supposed  to have acquire 
 the material constituting the ring component by capturing the gas during tidal interaction
 with  neighbor galaxies.
 The subsequent gravitational interaction combined with the 
 gaseous dissipation then  
 spreads the captured gas   
 and  forms  the polar-rings around the host galaxy.
 One of the promising models along  this orthodox scenario is the `preferred plane model' in which
 the differential precession of the rings and the gaseous dissipation cooperate to play
 a vital role in leading the acquired gas to settle into the stable polar orbit and finally to 
 form the polar-rings (Tohline \& Osterbrock 1982; Durisen et al. 1983; Schweizer et al 1983). 
 A number of numerical simulations have already
confirmed 
in what physical conditions the polar-rings are more likely to form  
and continue to exist for a relatively longer time-scale
(Habe \& Ikeuchi 1988; Christodoulou et al. 1992; Katz \& Rix 1992). 
Indeed these previously proposed models
 have provided a potential success
 in reproducing the polar-rings in S0 galaxies, however, these seem to be incapable of giving 
 sufficiently conclusive and persuasive answers 
 to the following seven  questions on the origin of the polar-ring
galaxies (Sackett 1991; Whitmore 1991; Arnaboldi et al. 1997; 
Galletta et al. 1997):
(1) Why are nearly all the central host galaxies morphologically classified as S0 ?
 (2) Why are  polar-ring galaxies so rare among S0 galaxies ? (3) Why do some polar-ring galaxies
 have a narrow  ring and some have annuli ? (4) Why are the mass and angular momentum
 of the ring component comparable or sometimes larger than  those of the host ? 
 (5) Why are the rings so `polar' ? (6) Why some polar-rings have considerably peculiar
 morphology such as helical and double-ringed shapes ?
 (7) Why is there an appreciably greater
 amount of interstellar gas in the polar-ring 
 component ?
 In particular, (1),  (4), and (7)
 could not be explained  simply by the previous theoretical  models,
 implying  either that more elaborated and sophisticated models along 
 the above orthodox  scenario should be considered 
or that the alternative model should be proposed for the 
explanation of the above questions.

 The purpose of this paper is to explore the origan of polar-ring galaxies and to propound a new mechanism
which more naturally  and reasonably 
explains the aforementioned observational trends of polar-ring galaxies.
In the present study, we consider that the dissipative galaxy merging between two disks 
is a promising mechanism that quite reasonably answers  the above seven questions.
Therefore we investigate how the dissipative galaxy merging transforms two disks into
one early-type S0 galaxy with polar-ring.
Furthermore, we investigate how the orbit configuration of galaxy merging and initial mass ratio
of the two progenitor disk galaxies can affect the morphology of polar-rings developed after galaxy 
merging.
In this paper, 
the galaxy merging with specific 
orbit configurations and sufficient  amount of gaseous dissipation
is demonstrated to play a vital role in forming both the cental S0-like host and the surrounding
polar-ring component in polar-ring galaxies.
This paper is an extended version of Bekki (1997) in which the basic mechanism of polar-ring
S0 galaxy formation in galaxy mergers is briefly summarized.
The layout of this paper is as follows.  \S 2 describes numerical models for dissipative galaxy merging.
\S 3 gives the results obtained in the present study.
In \S 4, we mainly discuss whether or not  the model proposed in the present paper can 
become a new promising model which naturally and reasonably 
explains the observational properties of polar-ring
S0 galaxies. 

\section{Model}
\subsection{Structural  and kinematical properties of initial  disks}

 We construct  models of galaxy mergers between gas-rich late-type
 disk galaxies by using Fall-Efstathiou model (Fall \& Efstathiou 1980). 
 In the present merge model, a galaxy intruding from the polar axis of the other
 galaxy in a merger is referred to as the `intruder' whereas the other is 
 referred to as the `victim'.
 In the present model, the dynamical and kinematical properties of 
 the victim 
 are set to be exactly the same as those of the standard model described below whereas 
 those of the intruder with  a given mass ($m_{2}$)
 and size ($r_{2}$) are determined by rescaling those of the
 standard model.
 Both the density profile  of the disk and the rotational curve profile  are assumed
 to be self-similar  between the two galaxies and these two galaxies
 satisfy the Freeman's luminosity-size relation (Freeman 1970). 
 For example, if we set the intruder mass to be 2.0, the size of the intruder
 is automatically set to be 1.41.

 In the  standard model (corresponding to the victim model),
 the total mass ($M_{\rm d}$) and the size ($R_{\rm d}$) of a progenitor disk  are 
 set to be 1.0 and 1.0, respectively. 
 From now on, all the mass and length are measured in units of
  $M_{\rm d}$ and  $R_{\rm d}$, respectively, unless specified. 
  Velocity and time are 
  measured in units of $v$ = $ (GM_{\rm d}/R_{\rm d})^{1/2}$ and
  $t_{\rm dyn}$ = $(R_{\rm d}^{3}/GM_{\rm d})^{1/2}$, respectively,
  where $G$ is the gravitational constant and assumed to be 1.0
  in the present study. 
  If we adopt $M_{\rm d}$ = 1.0 $\times$ $10^{10}$ $ \rm M_{\odot}$ and
  $R_{\rm d}$ = 10.0 kpc as a fiducial value 
  then $v$ = 6.53 $\times$
  $10$ km/s  and  $t_{\rm dyn}$ = 1.49 $\times$ $10^{8}$ yr,
  respectively.
  These fiducial values are appreciably  smaller than  those adopted in Bekki (1997),
  principally because we here intend to discuss  the formation of polar-rings in
  less luminous galaxies prevalent
  among polar-ring galaxies.
  In the present model, the rotation curve becomes nearly flat
  at  0.35  $R_{\rm d}$  with the maximum rotational velocity 2.22 in
  our units, corresponding to  total halo  mass within $R_{\rm d}$ equal to 3.58 
  in our units.
  The radial ($R$) and vertical ($Z$) density profile 
  of a  disk are  assumed to be
  proportional to $\exp (R/R_{0}) $ with scale length $R_{0}$ = 0.2
  and to  ${\rm sech}^2 (Z/Z_{0})$ with scale length $Z_{0}$ = 0.04
  in our units,
  respectively.
  The  velocity dispersion of dark  halo particles at a given point
  is set to be isotropic and given
  by the  virial theorem.
  In addition to a  rotational velocity set  by the gravitational
  field of disk and halo components, the initial radial and azimuthal velocity
  dispersions are assigned  to disk components according
  to the epicyclic theory with Toomre's parameter, $Q$ (Binney \& Tremaine 1987), 
  equal to 1.0.
  The vertical velocity dispersion at given radius 
  are set to be 0.5 times as large as
  the radial velocity dispersion at that point.

  The collisional and dissipative nature 
  of the interstellar medium are modeled by the sticky particle method (Schwarz 1981).
  It should be emphasized here that this discrete cloud model can at best represent
     the $real$ interstellar medium of galaxies  in a schematic way.
     As is modeled by McKee \& Ostriker (1977),
     the interstellar medium can be considered to be
      composed mainly of `hot', `warm', and `cool'
      gas, each of which mutually
      interacts hydrodynamically
      in a rather  complicated way.
      Actually, these considerably complicated nature of
      interstellar medium in  disk galaxies would not be
      so simply modeled by the `sticky
      particle' method in which gaseous dissipation is modeled by ad hoc
      cloud-cloud collision: Any existing numerical method probably could
      not model the $real$ interstellar medium in an admittedly proper
      way.
      In the present study, as a compromise,
      we only try to address some important aspects of hydrodynamical
      interaction between interstellar medium in disk galaxies and in
      dissipative mergers.
      More elaborated numerical modeling for real interstellar medium
      would be  necessary for
      our further understanding of dynamical evolution
      in dissipative galaxy mergers.
  The size  of the clouds relative to the 
  disk size is set to be 7.5 $\times 10^{-3}$ in our units 
  in the present simulations. 
     The corresponding size and mass of each cloud in the present study,
     are 75pc and $10^{5}$ $\rm M_{\odot}$, respectively.
  The radial and tangential restitution coefficient for cloud-cloud
  collisions are
  set to be 1.0 and
  0.0, respectively.
  In all the simulations, the initial gas mass fraction represented
  by $M_{\rm g}$ is set to be 0.0 for 
  the intruder and 0.2 for the victim unless specified. 
  The reason for this initial condition  is 
  principally that  
  we do not believe that interaction between two gas disks is
  essential for polar-ring formation in the present model, and
  it introduce a great deal of complexity.
  The adopted value of 0.2 is typical one for gas-rich 
  spiral galaxies (e.g., Roberts \& Haynes 1994).
  Total particle numbers used in the standard model (for the victim galaxy) 
  are 10000 for halo component,
  10000 for stellar disk one, and 20000 for gas one.
  For the intruder, the total particle numbers depend on the $m_{2}$ in such a 
  way that the numbers are linearly proportional to the $m_{2}$.
  For example, 
  total particle numbers  
  are 20000 for halo component and 
  20000 for stellar disk one, in the model with $m_{2}$ = 2.0.

\placetable{tbl-1}
\placefigure{fig-1}
\placefigure{fig-2}
\placefigure{fig-3}

\subsection{Star formation}
 We incorporate physical  processes of  star formation  into  the
 present  model in a  more idealized manner.
 We consider only  the  conversion of gaseous component
 to stellar one  and do not include 
 here  other important effects of 
 star formation such as
 effects of thermal and dynamical heating by   type II  supernova  on the 
 dynamics in this preliminary stage.
 In the present study, we will use the term `star formation'
 to refer to the process of gas consumption. 
A new stellar particle (collisionless particle, referred to as `new stars' or
as `new stellar component')
is  created at the position of original gas particle according to the algorism 
described below.
    We adopt the Schmidt law (Schmidt 1959) 
    with exponent $\gamma$ = 2.0
    as the controlling
    parameter of the rate of star formation.
    The amount of gas 
    consumed by star formation for each gas particle
    in each time step, 
    $\dot{M_{\rm g}}$, 
is given as
      $ \dot{M_{\rm g}} \propto  
 {(\rho_{\rm g}/{\rho_{0}})}^{\gamma - 1.0} $
    where $\rho_{\rm g}$ and $\rho_{0}$
    are the gas density around each gas particle and
    the mean gas density at 0.48 radius  of 
    an initial disk, respectively.
     The coefficient of the Schmidt law is set to be the same for
     the whole region of galaxies in the present study.
 As is described this, the `star formation' in this preliminary
study only means
the formation of collisionless particles and does not literally mean
the actual and realistic series of star formation
 such as successive fragmentation of gas clumps
 and the resultant formation of
 very dense molecule core.
This modeling for star formation is rather simplified so that
we can only address some important aspects of the roles of
star formation in  the formation of polar-ring galaxies. 
However, we believe that since the main point of the present study
is only the qualitative behavior of the star formation in
the formation of polar-rings,
even the rather simple model adopted in this study makes it possible to
grasp some essential ingredients of the roles of `real' star formation.
More extensive studies on this subject will be done in our future papers
by using more elaborated model for star formation.

\subsection{Orbit configuration of galaxy merging}
    In all the simulations, the orbit of the two disks  is set to be
    initially in the $xy$ plane and the distance between
    the center of mass of the two disks is  set to be initially
    4.0 in our units.
    The initial spin axis of the victim  is set to be in a $xz$ plane 
    and specified by 
    a parameter $\theta_{1}$ 
    which describes the angle between the $z$ axis and the spin 
    angular momentum vector of the victim disk.
    The initial spin of the intruder is set to be exactly parallel to $z$ axis
    for all models.  
    Two different types  of orbit  are adopted in
the present study.
One is that 
the two disks are set to move initially on a $x$ axis with absolute magnitude of
relative velocity  $V_{\rm  rel}$ (``polar-axis collision'').
This type of initial orbit configuration with nearly zero orbital
angular momentum is frequently adopted in the studies
of the formation of collisional galactic rings  (e.g., Hernquist \& Weil 1993;
Appleton \& Struck-Marcell 1996;
Gerber, Lamb, \& Balsara 1996).
The other is that 
the two disks are  set to
merge with  parabolic
encounter
and with a given the pericenter distance, $r_{\rm p}$  (``parabolic collision''). 
This type of initial orbit configuration with an appreciable amount
of orbital angular momentum in a bound system is adopted in the studies of 
elliptical galaxy formation in galaxy mergers (e.g., Barnes 1992).
The main reason for our adopting these two types
of  rather simple orbit configuration is that we intend to 
elucidate more clearly the formation mechanism of  polar-ring galaxies.
Formation of polar-ring galaxies in a more complicated and less idealized situation
of galaxy merging will be investigated in our future papers.

   In what follows, we investigate how a  polar-ring galaxy is developed during galaxy merging
   between two disks and clarify the formation mechanism of the polar-rings.
Furthermore, we vary the initial mass ratio of two progenitor galaxies ($m_{2}$) and thereby investigate 
how the initial mass ratio controls the final morphology of the developed polar-ring
galaxies in dissipative galaxy mergers.
   All the simulations have been carried out on the GRAPE board (Sugimoto et al. 1990)
   at Astronomical institute of Tohoku university.
   The  gravitational softening parameter is  fixed at 0.03  
   in all the simulations. 
   Considering the particle number of the present study (less than 100000)
   and the time-scale of 
   the numerical calculation (an order of  ten dynamical time),
   the two body relaxation due to the finite particle number
   (and due to the adoption of the softening length) has negligible dynamical effects 
   on the relatively global dynamics investigated in the present merger model.
   The time integration of the equation of motion
   is performed by using 2-order
   leap-frog method. 
$PO$ and $PA$ in the second  column of Table 1
specify the type of orbit configuration of galaxy merging:
$PO$ and $PA$ represent the polar-axis collision and parabolic one,
respectively. 
  The values of model parameters, $m_{2}$, $r_{2}$, $M_{\rm g}$,
$\theta_{1}$, $V_{\rm rel}$, and $r_{\rm p}$
   are  summarized in Table 1 for 17 models (Model 1 $\sim $ 17).
Characteristic of polar-ring morphology developed after galaxy merging are described
in  the ninth column of Table 1 for each model. 
For example, the `double ring' means that the developed polar-ring morphology
$looks$ $like$ double-ring.

\placefigure{fig-4}
\placefigure{fig-5}
\placefigure{fig-6}
\placefigure{fig-7}
\placefigure{fig-8}

\section{Result}
In this section, we first observe how a S0 galaxy with polar-rings is formed in a dissipative merger
with  particular orbit configuration (\S 3.1).
Next we describe how the initial mass ratio of two progenitor
disk galaxies ($m_{2}$) 
controls the final morphology of the developed polar-rings (\S 3.2).

\subsection{Dynamics  of polar-ring S0 formation}
 First, we present the results of Model 1 which shows a typical behavior of polar-ring formation 
in the present study.
S0 galaxies with narrow polar-rings are found to be more likely to form in this model
with the initial mass ratio
of two progenitor disks,  $m_{2}$, equal to  2.0, as is shown in  Figure 1.
The time evolution of merger orbit projected onto $x$-axis for Model 1 is given in
Figure 2.
Time evolution of gas mass for this model is given in Figure 3.
In this model the intruder galaxy is found to be
inevitably transformed  into the central S0-like component whereas the victim is found to be dramatically
transformed into a narrow polar-ring, as is described below.

\subsubsection{Mechanism of polar-ring formation}
Figure  4, 5, 6,  and 7 describe the morphological evolution of each component in
each of two galaxies.
As is shown in these figures,
as the intruder pierces  the victim and shoots through it,
the initially thin stellar disk of the intruder become thicker and thicker owing to strong dynamical 
heating during the violent gravitational interaction between these two galaxies (the
time, $T$, equal to 2.0 in our units).
While the intruder is then pulled gravitationally back to pierce again the victim ($T$=4.0),
the stellar disk of the intruder suffers from violent relaxation to form 
more disky S0-like spheroid ($T$=8.0).
The density profile of the intruder at $T$ = 14.0 shows clear
deviation from initial exponential profile looking more like 
the so-called $R^{1/4}$ law,
and furthermore its kinematics within  the central host shows 
an appreciable amount of global
rotation (See Figure 8.). 
These results clearly demonstrate that the 
formation of S0-like hosts in  polar-ring galaxies is
closely associated with the violent dynamical interaction
of the two merging galaxies  and furthermore that the formation
of S0-like host (which is the intruder in the present model)
is inevitable in the formation of polar-ring galaxies
owing to the stronger gravitational interaction of galaxy merging.

As is shown in Figure  4, 5, 6, and 7, 
the victim disk, on the other hand, is finally transformed into
a narrow polar-ring surrounding the central spheroid 
mainly composed of   the stellar disk of the intruder. 
As the intruder pushes the central part of the victim out and simultaneously excites the outwardly
propagating density wave in the victim, the stellar component of the victim is dynamically relaxed
to form the diffuse spheroidal component  in  the merger remnant because 
of violent relaxation of galaxy merging.
The gaseous component of the victim
forms the compressed gas layer with higher density owing to 
the enhanced cloud-cloud collisions and the resultant gaseous dissipation. 
Subsequent star formation and further gaseous dissipation at 
the compressed gas layer dramatically
transforms the victim disk into a narrow 
polar-ring composed of gas and new stars.
Morphology of the gaseous polar-ring projected 
onto the $xz$ plane looks more like `integral-sign' 
galactic warp at $T$ = 10.0 and `banana' one at $T$ = 12.0,
implying that warps frequently observed in polar-ring galaxies are closely
associated with the gas and stellar dynamics of the merging victim galaxy. 
The victim at $T$ = 14.0
shows mass distribution clearly  deviating  from exponential
profile and does not have any remarkable global rotation
along the major axis of the central host (See Figure 8.).
As is shown in Figure 9, the mass distribution of new stars formed  after
$T$ = 8.0 
seems to be more disky  whereas that of new stars  
formed before   $T$ = 8.0 seems to be more extended and spherical.
This result reflects the fact that an appreciable amount of new stars are produced in
the shocked and ring-like gas layer during the galaxy merging.
This new stellar component formed in the relatively later phase of galaxy merging
can be observed as the stellar polar-ring in polar-ring galaxies.
These results quite naturally explain 
why some polar-ring galaxies have an appreciable amount
of interstellar gas in the polar-rings: This is simply because the polar-ring
component is originally a
gas-rich galactic   disk. 
What we should stress furthermore  is that in this Figure 9, 
 a discernable amount of interstellar gas and new stars
  is fueled to the central part of the merger remnant.
  This result is consistent with the observational evidences that
  some polar-ring galaxies actually show the pronounced activity
  (e.g., post star-bursts and LINER activity) in their central part
  (e.g., Schechter et al. 1984; Reshetnikov \& Combes 1994).
  This result also implies that such polar-ring galaxies as showing
  the central activity are examples of polar-ring S0s formed by merger 
  events.

    As is described above, the successive radially propagating wave excited
    by the intruder is found to play an important role in the earlier  
    stage of  polar-ring formation.
    Moreover, the gaseous dissipation and the subsequent star formation in
    the compressed gaseous region in the wave are found to damp
    the radial oscillation of the $wave$ and then to change the $wave$ into
    the stational polar-ring $material$ in the later stage of polar-ring formation.
    Such damping of the radial oscillation owing to gaseous dissipation
    in galaxy merging is described
    by  Figure 10 in which the same analysis as that of Athanassoula,
    Puerari, Bosma (1997) is adopted.
    This two-fold polar-ring formation is a unique physical process of dissipative
    galaxy mergers  with specific  orbital  configurations,
    thus provide an alternative explanation for the polar-ring formation.
    The developed polar-ring component is found to show discernable precession
    even after a  merger remnant reaches  the  nearly dynamical equilibrium state,
    reflecting the fact that the formation of the polar-ring is a result of the 
    past violent gravitational interaction between two disk galaxies.
    This precession is one of characteristics
    of polar-rings developed in the present merger model for polar-ring formation,
    as is described in detail later.

\placefigure{fig-9}
\placefigure{fig-10}

Thus  a central S0-like host galaxy with a narrow polar ring is formed by dissipative galaxy merging
with  specific  orbital  configurations
within only $\sim$ 12 $t_{\rm dyn}$ corresponding to 
$\sim$ 1.8 $\times$ $10^{9}$ yr. 
These results demonstrate that strong dynamical effect  on the central part of the victim disk
is indispensable for the formation of polar-rings in this picture,
which is quite different from the previously
proposed models,  such as the preferred plane models.
Moreover the time-scale of polar-ring formation in the present merger model,
which is typically $\sim$ 10 dynamical time
of the system, is relatively short compared to  that in the preferred plane one. 
The present merger model for polar-ring formation accordingly seems to
be a new and promising candidate theory, which can give reasonable and plausible explanations 
for a number of fundamental observational properties of polar-ring galaxies.
However it is safe for us to say that it is not certain whether or not the present
merger model is an $unique$ mechanism for polar-ring formation: There are a number of
observational indications 
that in a certain interacting galaxy,  the  polar-ring $seems$ to be 
about to form according to 
the preferred plane scenario (e.g. Reshtnikov, Hagen-Thorn, \& Yakovlera 1996).
Observability of polar-ring S0s formed by dissipative 
galaxy merging is discussed in details
in the \S 4. 

\subsubsection{Physical conditions required for polar-ring S0 formation}
 We here describe  under what physical conditions 
 S0 galaxies with polar-rings are more likely 
to form in dissipative galaxy merging.
We here observe the results of four comparative experiments of
models, Model 2, 3, 4, and 5, which are designed to elucidate
more clearly the essential parameters for the formation of polar-rings.
Figure 11 gives a collection of the final morphology
of the three models (Model 3, 4,
and 5).
We  find that neither the model with less 
inclined victim  disk (${\theta}_{1}$ = 30.0, Model 3) nor the model with an appreciably
larger amount of initial orbital angular momentum (Model 4) show
any remarkable polar-ring component.
This result suggests that both the more inclined victim disk and a relatively smaller
amount of orbital angular momentum are required for the formation of polar-rings in
the present merger model (See Figure 11.).
This  result of Model 3 
explains why the polar-rings are so polar (e.g., Whitmore 1991): This is because if the two 
progenitor disks are not so highly orthogonal with each other, the merger remnant just
becomes a morphologically normal S0 galaxy.
The result of Model 4, the orbit configuration of which
has been  investigated more extensively by several authors in the context of E/S0
galaxy formation,
naturally explains why the polar-ring E/S0 galaxies are so rare among E/S0 galaxies:
This is because a specific kind of orbit configuration of galaxy merging is 
required for the formation of polar-ring E/S0 galaxies.
We also find
that the model with no gas particle (i.e, without gaseous dissipation), Model 5, does
not show any ring-like component 
surrounding the central spheroid developed after galaxy merging (See Figure 11.).
This result suggests the following important role of gaseous dissipation in 
 the polar-ring formation:
 Gaseous dissipation can remove random kinetic energy of the gaseous component
  in the victim disk, which is caused by violent gravitational interaction between
  two disks, and accordingly it enables some of the gas component to settle again in 
  well-ordered disky
  configuration.
This result of Model 5 naturally explains why  polar-ring galaxies are less
massive or luminous on average: 
This is principally because less luminous late-type spirals
generally have an  appreciably larger amount of interstellar gas (e.g., Roberts \& Haynes 
1994), which is indispensable for polar-ring formation.

These comparative experiments clearly demonstrate that both a minimum
 amount of gaseous dissipation 
and a specific  kind of orbit configuration are required for the formation of polar-rings,
which naturally explains why the polar-ring galaxies are so rare among S0 galaxies.
In this model with $m_{2}$ = 2.0, the polar-ring morphology depends weakly on the relative 
velocity of two merger precursor disk: As the relative velocity becomes larger (e.g,
in the model with $V_{\rm rel}$ = 1.5, Model 6),
the polar-ring component looks more like a disturbed dynamical system. 

\placefigure{fig-11}
\placefigure{fig-12}
\placefigure{fig-13}

\subsection{Dependence on initial mass ratio of two progenitor disks}

 In this subsection, we mainly describe how the initial mass ratio of two progenitor
galaxies affects the final morphology of the central S0-like spheroid and the surrounding
ring component developed after galaxy merging.
Figures 12 and 13 summarize the final mass distribution for models with $m_{2}$ =
1.5, 1.0, 0.7, 0.5, 0.3 and 0.1
(corresponding to Model 7, 8, 9, 13, 16, and 17, respectively)
in which orbit configuration of galaxy merging is exactly the same as that of Model 1
but the initial mass ratio of progenitor disks
($m_{2}$) is different from that of Model 1. These figures clearly indicate the 
$m_{2}$ dependence of the final morphology of polar-ring galaxies formed by dissipative galaxy
merging, as is described later.
In the following,
for each model with different $m_{2}$, the dependence of
structure and kinematics of the developed polar-ring galaxies 
on the relative velocity of two merger
precursor galaxies are given only if the results of this dependence are considered to be
remarkably important.
 
\subsubsection{$m_{2}$ = 1.5}

Like  Model 1, the model with $m_{2}$ = 1.5 
(Model 7) shows both the central disky S0 host
and the surrounding thin and narrow polar-ring component (See Figure 12.).
The essential physical processes of polar-ring formation are nearly the same
as those in Model 1.
The central S0 in this model looks more    puffed out compared with that of Model 1,
probably because the intruder suffers from 
the stronger dynamical interaction with the victim disk.
The axis of intrinsic angular momentum  of the developed S0 galaxy is more
appreciably inclined compared with initial intrinsic spin axis of the intruder.
These results suggests that both morphology of the central S0 host and the inclination
angle between polar-rings and the central host in a polar-ring S0 galaxy can 
reflect the initial mass ratio ($m_{2}$) of two merger precursor disks,
which furthermore suggests that the strength of dynamical interaction in galaxy
merging is an important factor for structure and kinematics of merger remnants.
In the present study,
S0 galaxies with a narrow polar-ring (such as ESO 415 - G 26) are found to be reproduced
more successfully in the models with $m_{2}$ larger than 1.5. 

\subsubsection{$m_{2}$ = 1.0}
 For the model with $m_{2}$ = 1.0 (corresponding to the so-called
 `major merger', Model 8), both the intruder disk and the victim one 
 suffer from the stronger
 violent dynamical relaxation of galaxy merging, principally because the
 total mass of the intruder is exactly the same as that of the victim.
 Consequently, the central spheroidal galaxy developed after galaxy merging
 looks more like an elliptical galaxy. The polar-ring component,
 on the other hand,
 looks more strongly disturbed
 than those of Model 1 and 7 probably because of the stronger 
 dynamical relaxation of  galaxy merging. 
 For this model (and the models with 0.5 $\leq$ $m_{2}$ $\leq$ 1.0), 
 the dynamical interaction between the developed central
 spheroid and the polar-ring component is very significant  so that
 the polar-ring shows discernable precession around the central host after galaxy merging. 
 Considering the relatively smaller probability of major merging,
 polar-ring galaxies developed in major galaxy mergers like this Model
 8 could be considerably rare,
 which can give a natural explanation for the extremely rare existence 
 of elliptical galaxies with polar-rings (such as AM 2020-5050).

\placefigure{fig-14}

\subsubsection{$m_{2}$ = 0.7}
 Essential physical processes of polar-ring S0 formation in the model
 with $m_{2}$ = 0.7 (Model 9)  are  basically the same as
 those of Model 1, as is described in Bekki (1997).
 In this model,
the transient morphology of the ongoing merger at the time when the intruder
pierces the victim disk is strikingly similar to the 
morphology of the two disks in NGC 660 (e.g., van Driel et al. 1995).
This result suggests that galaxies like  NGC 660 are characteristic  
and probably rare examples of polar-ring galaxies formed by dissipative galaxy 
merging.
Even for this model,
both density profile of the intruder and that of the victim 
ultimately  show clear
deviation from their initial exponential profile looking more like $R^{1/4}$ law,
and furthermore kinematical property of the central host shows appreciable amount of global
rotation.
What is remarkable in this model is that  the ring morphology in the merger remnant depends
more strongly on the relative velocity of the two galaxies.
It is found that the larger the initial relative velocity is, the more peculiar the
morphology of the developed polar-ring is (Model 10 and 11): For models with the
relative velocity is larger than 1.5 but less than the escape velocity of the system,
some merger models show double-rings or other peculiar
structure especially in the gaseous component
of the merger remnant.
Furthermore, as has already been described in Bekki (1997),  polar-rings
do not form
but rather  transient `Cartwheel'-like rings  form in the victim disk 
if the relative velocity of the two merger precursor exceeds the escape
velocity of the system (Model 12).
In this model  (Model 12), the
intruder galaxy completely having escaped from the gravitational trap of the victim
is transformed into S0-like galaxy after the galaxy interaction.
These  results imply that only the initial relative velocity of merger precursors
is the difference in the formation process between the apparently different two types of 
galactic rings (that is, polar-rings and collisional `Cartwheel'-like rings).

\subsubsection{$m_{2}$ = 0.5}
  It is found that if the $m_{2}$ is approximately equal to  0.5, 
  the stellar disk of the victim
  galaxy is not so completely destroyed by the intruder galaxy during galaxy merging
  primarily because the intruder does not  dynamically disturb  the victim
  disk strongly enough  to cause violent dynamical relaxation.
  This incomplete destruction of the victim disk results in peculiar morphological
  evolution of polar-rings surrounding the central early-type galaxy in a galaxy merger.
  For the model with $m_{2}$ = 0.5 (Model 13), the gaseous (and new stellar)
  component in the merger remnant show precession  even after the completion of galaxy
  merging,
  principally because the gaseous component and the stellar one in the merger remnant
  mutually interact with each other even after galaxy merging.
  Consequently, the long-term dynamical evolution of gaseous component in this model is 
  remarkably different from that of other models with $m_{2}$ =  2.0 and 1.5  (Model 1 and 7).
  As is shown in  Figure  14, the gaseous polar-ring developed  after galaxy merging
  finally $looks$ $like$  a double-ringed system owing to mutual dynamical
  interaction with the stellar
  component of the merger remnant (at $T$ = 26.0).
  We also find that this peculiar double-ringed morphology can be observed in the
  models with different initial collisional parameters  of galaxy mergers 
  (Model 14 and 15). 
 These results suggest that the formation of S0 galaxies with morphologically 
peculiar polar-rings such as helical rings (e.g., NGC 2685) and double rings
(e.g., ESO 474-G 26) is  closely associated with the later dynamical interaction
between the stellar merger remnant and 
the developed polar-rings.
 These results furthermore imply that the mass ratio of merger precursor galaxies 
 required for the formation of double-ringed early-type galaxies should be approximately
 0.5.
Future numerical studies with high resolution  and with
more sophisticated gas dynamics and star formation
would enable us to reproduce more successfully
the S0 galaxies with peculiar polar-rings which are 
strikingly similar to NGC 2685 and  ESO 474-G 26 both in kinematics and mass distribution.

\placefigure{fig-15}
\placefigure{fig-16}
\placefigure{fig-17}
\placefigure{fig-18}
\placefigure{fig-19}

\subsubsection{$m_{2}$=0.3}
We find that if the $m_{2}$ is approximately less than 0.3, the victim galaxy is not 
substantially disrupted 
by the intruder  and only  suffers from  dynamical thickening of the disk,
whereas the intruder  is completely destroyed by the strong disturbance.
Figure  15, 16, 17   and 18
describe how a   central host galaxy with polar-rings which $looking$ $more$
$like$ annular rings is formed during dissipative galaxy merging in the model with $m_{2}$
= 0.3 (Model 16).
As is shown in these figures, 
the intruder galaxy with smaller mass is completely destroyed 
during violent relaxation of galaxy merging to
form a less centrally concentrated spheroid in the merger remnant.
As a result of this, the intruder does  not  dynamically   damage the victim strongly enough
to trigger the violent relaxation of the victim stellar disk.
Accordingly, 
the stellar disk in the victim is not destroyed  completely to keep the disk configuration
even after galaxy merging.
In the gaseous disk of the victim, outwardly propagating gaseous waves, which are excited by 
the intruder galaxy during galaxy merging, are  transformed into the annular gaseous rings
at $T$ = 6.0 and finally to non-axisymmetric  structure at $T$ =12.0.
In total,  more than fifty 
percent of initial gas mass is converted to new stars within 14.0 $t_{\rm dyn}$ 
and most of the new stars are located in the victim disk even after galaxy merging
in this model.
The warp of gaseous disk shows  variously different morphology at different time,
which suggests that morphological types of warps observed 
in polar-rings can reflect the time dependent gas dynamics in the polar-rings (or
victim disk).
As is shown in Figure 19, the radial mass distribution of
the intruder shows clear deviations from initial exponential profiles
owing to  the stronger dynamical relaxation in the  
galaxy merging.
The developed disk  with morphologically peculiar gaseous component might be actually
observed as annular polar-rings  in the nearly edge-on view of the galaxy.

Thus a smaller spheroidal
galaxy with a larger and morphologically peculiar gaseous disk and a stellar
disk are found to be formed, which 
$looks$ $like$ the edge-on view of  a S0 galaxy with $transient$  annular
polar-rings such as NGC 4650a.
In particular, the face-on view of the victim disk in Figure 17 at $T$ = 4.0
could be  similar to that of annular polar-ring galaxies. 
The annular-ring feature in Figure 17 at $T$ = 4.0 furthermore
reminds us of the Hoag's objects (Schweizer et al. 1987), which
suggests a close physical connection between annular polar-ring galaxies
and the Hoag's objects.
Appleton \& Struck-Marcell (1996) describe the details of the possible 
mechanism for the formation of the Hoag's objects in terms of 
epicyclic interference patterns resulting from multiple encounters.
It is not clear, at least in the present study, whether 
the Hoag's objects are actually face-on views of annular polar-ring S0s
or forming collisional rings such as a `Cartwheel' ring. 
Furthermore there are no extensive studies investigating
whether or not 
such epicyclic interference patterns as are described
in Appleton \& Struck-Marcell (1996) can provide a 
mechanism for the formation of annular polar-rings.
It should be accordingly
addressed in our future papers whether the mechanism 
for the formation of Hoag's objects is essentially the same as that
for the polar-ring S0 formation.
Anyway the 
results derived  in the model
with $m_{2}$ = 0.3
provide a clue to the understanding of the origin of the appreciably larger amount
of mass and angular momentum in polar-rings observed in NGC 4650a:
The larger `polar-ring' component in NGC 4650a was once  a more massive gas-rich
disk galaxy which had finally
transformed into `polar-rings' because of the past galaxy merging with a less massive galaxy. 
The  results of this Model 1 and those of Model 16 imply  that both narrow polar-rings
and annular  ones, which are considered to be
the two basic morphological types of polar-ring galaxies,
are
originally   disk galaxies before they  manifest themselves  
as polar-rings  and that the dichotomy between the narrow rings and annular ones
can be understood in terms of the difference in the initial mass ratio of two merger
progenitor
disks in galaxy merging.

What we should stress here is that we have only succeeded in reproducing a central  galaxy whose
edge-on view is apparently similar to the S0 galaxy with annular polar-rings.
Actually the central galaxy developed in the model with $m_{2}$ = 0.3
does not show so large global rotation as is observed in 
annular polar-ring galaxies (e.g., NGC 4650a),
as is shown in Figure 19.
Furthermore it is clear from the mass distribution
of the merger remnant (Figure 19) that 
the central part of the merger remnant is not completely cut out.
That is, we appear to have failed to reproduce  an annular polar-ring
galaxy 
with the central part of the ring component completely cut out
and with the central host showing a relatively larger amount of
global rotation.
This apparent failure implies either that the galactic model adopted in the present study,
including the star formation law, gas physics, and the assumption of Freeman's law
(that is, self-similar disks), 
are not so appropriate
or that 
S0 galaxies with annular
polar-rings cannot be formed by galaxy merging.
Considering the relatively smaller parameter space investigated in the present 
study and the present rather idealized gas dynamics and star formation of dissipative 
galaxy mergers, 
it is safe for us to say that 
it is not clear, at least in this preliminary stage of the present study,
whether our future elaborated numerical models can simulate the formation of a S0 galaxy 
with annular polar-rings.
The problems concerning the formation of annular polar-rings and the promising ways to
remove these are discussed in detail in the section of discussion (\S 4).

\subsubsection{$m_{2}$=0.1}

It is found that if the initial mass ratio is less than 0.1, neither the central spindly S0-like
host nor the disk component with peculiar morphology are formed.
The main reason for this is  that  the 
intruder galaxy is completely destroyed during merging 
to form a considerably diffuse spheroid, which would
not be identified with a spindly S0 galaxy, and thus  can
not give the stronger dynamical impact to the victim disk.
This result suggests that in addition to the particular orbit configuration of galaxy
merging and gaseous dissipation, a certain range of mass ratio of merger precursor galaxies
(basically $m_{2}$ $\geq$ 0.3) is required for the formation of polar-ring galaxies.

\subsubsection{Brief summary of  $m_{2}$ dependence}
 As is described above, a spiral  
 galaxy intruding from the polar axis of the victim  galaxy excites the outwardly propagating 
 density wave in the gaseous component of the victim.
 The subsequent  gaseous dissipation and star formation dramatically transform the victim
 into polar-rings.
 The intruder galaxy, on the other hand, is  inevitably transformed into a rapidly rotating and 
 spindly S0 galaxy  owing to the violent  gravitational interaction of galaxy merging.
 One of the advantages in the present merger
 model of polar-ring S0 formation is that depending on the $m_{2}$ (and $V_{\rm rel}$),
 variously different morphology of polar-rings in polar-ring galaxies can be reproduced:
 As the $m_{2}$ becomes smaller, the polar-ring(s) can change from a narrow ring to peculiar
 double rings, and to transient annular rings.
 This result reflects the fact that the structure and kinematics of merger remnants
 depend strongly on the strength of the dynamical interaction of galaxy merging
 (or the degree of violent relaxation), which is basically determined by the mass
 ratio of merger precursor disks.
 The present explanation for the origin of variously different polar-ring
 morphology in polar-ring S0 galaxies is difficult to be observationally confirmed,
 however, the present scenario for the formation of polar-ring galaxies
 seems to provide a clue to the thorough understanding of the formation and evolution
 of polar-ring galaxies.

\section{Discussion}

\subsection{Possible candidates of forming polar-ring galaxies}

 The best way to assess the validity of the present merger model
 of polar-ring S0 formation is to observe the polar-ring galaxies
 just forming in galaxy mergers.
 Considering the relatively short time-scale of polar-ring formation
 in the present model 
(within  a several dynamical times of the merger system, corresponding
to less than $10^{9}$ yr, depending on the initial mass of the system),
the observable number of such forming polar-ring galaxies 
is likely to be small.
There are   a few examples, however,
especially in the Atlas of Peculiar
Galaxies (Arp 1966) and
the catalog of polar-ring galaxies (Whitmore et al. 1990),
which provide us with valuable information on the formation
process of polar-ring galaxies.
NGC 660 is one of the most promising candidates, 
in which the central host is surprisingly classified
by  a late-type  galaxy (Sb or Sc) and the ring component has an
  appreciably larger  amount of HI gas.
This polar-ring galaxy is possibly  an ongoing galaxy merger between two late-type spirals 
with exponential disks and the central 
spiral host galaxy would be transformed into early-type galaxy within $\sim 10^{9}$ yr because of
continuing  violent relaxation in the  galaxy merging process. 
Furthermore this galaxy actually does not show 
noticeable  tidal tails, a fact of which  is consistent with the present 
result that 
the ongoing mergers  required for  the formation of polar-ring 
galaxies do not develop  tidal tails.
An another example is the ESO 199 - IG 12 (Schweizer et al. 1983;
Whitmore et al. 1990) 
in which the center of the polar-ring component seems  $not$ to coincide
with the center of the central host S0 galaxy and a number of tidal debris 
can be observed in the edge of the S0.
This galaxy 
strongly suggests that the polar-ring component is still developing 
in an ongoing galaxy merger.
NGC 5544 (and 5545) and NGC 6240  could be also promising candidates of
forming polar-ring galaxies in galaxy mergers.

All of these possible polar-ring galaxies in formation
contain valuable information about 
the formation and evolution histories of the galaxies.
Accordingly more extensive
observational studies 
which address in more detail
the kinematical and structural properties of these systems 
are very desirable.
The following three are important diagnostics for polar-ring
S0 galaxies in formation.
The first is  
the relative radial velocity of the two interacting galaxies (if any and if possible),
which is a key factor determining
whether the interacting galaxy becomes a cartwheel-like ring galaxy (if the
relative velocity  is larger than the escape speed of the system) 
or a polar-ring
one (if not), as is indicated by Bekki (1997). 
The second is to confirm wether or not there are diffusely dispersed stellar components
(e.g., long-lived stars) in ongoing mergers.
The present merger model predicts that in the formation of polar-rings,
a greater amount of the victim stellar components is dynamically
pushed out by the intruder
in a galaxy merger.
The third is to check wether the central  host in a possibly
forming polar-ring galaxy is oblate spheroid or prolate one.
In the present merger model, the central host is a still oblate disk galaxy 
when the polar-ring is developing in a ongoing merger.
It is doubtlessly worthwhile to investigate observationally
a number of physical properties of possible forming polar-ring galaxies 
along the above three diagnostics.

What we should furthermore stress is that if polar-ring S0 galaxies 
are formed in such a way that the present merger model predicts, the suitable site 
for observing  polar-ring S0s in formation is higher redshift universe.
This is primarily  because the smaller relative velocity required for
polar-ring formation in galaxy mergers is more likely in higher redshift universe
(where multiple galaxy encounters can happen more frequently
and these encounters can contribute to the formation of
galaxy mergers with such lower relative velocity) rather than in
the lower redshift universe (where a dynamical system including possible galaxy mergers
is fully virialized and thus has a larger velocity dispersion in the member galaxies). 
If galaxy mergers with smaller relative velocity are actually
prevalent in higher redshift (thus, if
most of polar-ring S0s are formed at higher redshift),
polar-ring components observed in the present epoch are `older'.
Although  more extensive observational studies including not only dynamical and
kinematical properties of polar-rings  but also chemical and photometric evolution
of polar-ring galaxies are indispensable for clarifying the formation epoch,
a growing number of recent
observational results seem to support the `older' polar-rings.
Eskridge \& Pogge (1997) reveal that abundance of  H II regions in polar-ring
galaxy NGC 2685
amounts to  0.8 $\sim$ 1.1 solar abundance, which means that chemical evolution
in the polar-ring proceeds for a longer time-scale (corresponding to the
chemical evolution time-scale of typical galactic disks).
Brocca, Bettoni, \& Galletta find (1997) that there is no remarkable environmental
difference between polar-ring S0s and normal galaxies,
which suggests that if polar-rings are formed in interacting/merging galaxies,
the preferred epoch of interaction/merging is not present but past.
Reshetnikov (1997) reveals that the detection rate of possible polar-ring galaxies
in the sample of Hubble Deep Field (HDF) is extremely higher  (0.7 percent 
among  the HDF galaxies)
than that of local universe (0.05 percent),
which implies that  polar-ring galaxies are more likely to form in higher redshift. 
These observational studies does not necessarily mean that the preferred epoch of polar-ring
galaxy formation is higher redshift universe, however, these study seem to
suggest that polar-rings in formation can be more easily observed in higher redshift universe.

\subsection{Connection with other morphological types of E/S0 galaxies} 
 Galaxy mergers are generally considered to transform two disk galaxies 
into one E/S0 galaxy with variously different morphology 
and kinematics (e.g., Barnes \& 
Hernquist 1992).
For example, Barnes (1992) demonstrates that elliptical galaxies
with the radial density profile of the so-called 
$R^{1/4}$ law and smaller specific angular momentum
are reproduced reasonably well in disk-disk mergers owing to the efficient
transfer of mass and angular momentum during the merging.
Bekki \& Shioya (1997a) suggest that both boxy and disky elliptical
galaxies are formed by star-forming and dissipative galaxy mergers, 
depending on the star formation history of merger precursors.
Multiple merging, which could be occurred in the group of
galaxies, are also demonstrated to produce normal
elliptical galaxies (Barnes 1989; Weil \& Hernquist 1996).
Furthermore, barred early-type galaxies can be reproduced in galaxy mergers 
with prograde orbit configuration (Mihos, Walker, \& Hernquist 1995,
Bekki \& Shioya 1997b) as well as in interacting galaxies (Noguchi 1987).
In the present study, polar-ring galaxies rather than the above `normal'
E/S0 galaxies are produced, implying a close physical relationship
between polar-ring galaxies and the above `normal' E/S0 galaxies.
Considering the physical conditions required for polar-rings in the 
present study,
it seems to be reasonable to claim that 
the physical conditions governing the difference in
structure and kinematics between these two apparently
different types of galaxies (polar-ring S0s and normal E/S0 galaxies)
are initial orbit configuration of galaxy 
merging and mass-ratio of two interacting/merging  galaxies: If two
gas-rich disk galaxies with unequal mass merge with each other  
with relative inclination of the two galaxies  highly perpendicular
with each other  
and with the relative collisional velocity relatively smaller (less
than escape velocity of the system), the merger remnant becomes
a polar-ring galaxy, otherwise they become a `normal' E/S0 galaxy.
Thus, the present study strongly suggests that `normal' E/S0 galaxies
and polar-ring galaxies are `relatives' in the sense that they
are all merger remnants with different initial conditions of galaxy
merging.

\subsection{Future study}

\subsubsection{Longevity and stability of polar-rings}
 Although it appears that the present merger model has succeeded in reproducing
both the central S0-like host and polar-ring component in polar-ring galaxies,
there still remains a number of issues we should address in order to confirm
how plausible and viable the merger model actually is.
Among these, 
  we  must first investigate how long  
 polar-rings developed via dissipative merger events
can remain dynamically stable and look more like `polar-rings'  after the merger remnant 
reaches the virial equilibrium.
Longevity and stability of polar-rings have been  demonstrated to depend  on the
characteristics of gaseous cooling and the degree of gaseous self-gravitation in
the ring component as well as on 
the three dimensional shapes of dark matter halo (e.g., Christodoulou et al. 1992;
Katz \& Rix 1992).
In the present merger model, especially
for the model with mass ratio of two disks approximately equal to 0.5 ($m_{2}$ = 0.5),
the developed polar-rings are  observed to show precession  even after 
the completion of galaxy merging owing to the dynamical interaction between the central host
galaxy and the ring component.
In this model, most of the materials in the victim disk is dynamically dispersed
owing to stronger tidal interaction between two disks, and thus self-gravity of
the victim disk is not important.
This result implies that in addition to the gaseous cooling and self-gravity, the dynamical
interaction between the central host and polar-rings is one of the key 
determinant for the
stability and longevity of the polar-rings.
In the present stage of this preliminary work, however,
it has not  been demonstrated  whether or not 
the developed polar-rings can survive 
for more than several tens of dynamical times of the system (corresponding roughly
to several $10^{9}$
yr) without
being destroyed or absorbed into the central host because of the later dynamical interaction
between these two components.
Thus, in our future study, we must
investigate how the later dynamical interaction between the central
host and the polar-ring component can determine the longevity and the 
long-term stability of the polar-ring
component,
in order to confirm the validity of the present merger model for
polar-ring(s) formation.
In particular, we intend to examine the dependence of the stability of polar-rings
on the initial mass ratio of progenitor two disk galaxies in galaxy mergers.
These studies would help us to predict how frequently we can observe the polar-ring galaxies among S0 galaxies
and furthermore give us more quantitative answers 
to the question as to why only  0.5 percent of the S0 galaxies actually show
remarkable polar-rings.

\subsubsection{Formation  of NGC 4650a}
 S0 galaxies with annular polar-rings (e.g., NGC 4650a) are considered to
 show global rotation in the central S0 and have 
 polar-rings with the central part completely cut out 
 (e.g., Schechter et al. 1984;
 Whitmore et al. 1990; Sackett et al. 1994). 
 Although 
 a S0 galaxy with narrow polar-rings (e.g., ESO 415 - G 26) 
 has been  demonstrated to be reproduced relatively successfully by dissipative galaxy
merging,  we  $appear$
to have failed to reproduce  exactly  such annular polar-ring galaxies.
Actually a self-gravitating disk with very peculiar edge-on morphology
observed in  models  with $m_{2}$ = 0.3, the transient feature of which can be
seen as annular polar-rings, does not have such a central hole as 
the observed annular polar-rings have.
One  interpretation of this apparent failure  is that 
the present merger model can 
$only$
form S0 galaxies with  narrow polar-rings (not with annular rings)
and thus that we should incorporate another important physical processes of
galaxy merging for the explanation of annular polar-ring galaxies.
The other possibility is that the apparently `annular' rings are not actually annular rings,
but  parts of galactic disks.
Recent observational studies 
on  a typical annular polar-ring galaxy, NGC 4650a, 
show that the morphology of the polar-ring component is  more like $spiral$ $arms$
rather than  annular rings (e.g., Arnaboldi et al. 1997).
This result suggests that the apparently `annular'
ring component is actually a self-gravitating `galactic disk' with spiral arm
morphology viewed from
a specific angle.
Furthermore, extensive observational study on the peculiar polar-ring galaxy NGC 660 shows
an exponential light
profile in the gas-rich polar-ring component, meaning that the polar-ring component is actually
a gas-rich spiral `galaxy' without  prominent  central part  (e.g., van
Driel et al 1995). 
These observational studies suggest that 
annular rings observed in polar-ring galaxies are actually not
the `rings'  but  normal galactic 
disks whose central density  becomes very  small
for  some unknown  reasons.
Further observational studies are desirable to  confirm   whether 
the central part of annular rings in real polar-ring galaxies are $completely$ 
cut out or the apparently
annular rings are in fact  a galactic disk with peculiar morphology.

\section{Conclusions}
  The present numerical study provides a new mechanism by which both the central S0-like host 
  and the ring component in a polar-ring galaxy are simultaneously formed.
  Although uncertainties of the numerical treatment of gas dynamics and star formation still remain,
  it appears that our model has succeeded in reproducing $some$
  polar-ring galaxies and explaining
  naturally a number of important observational properties of them.
  In the proposed model, the formation of polar-ring galaxies
  is essentially ascribed to 
  the details of dynamics of dissipative galaxy merging with  
  specific  orbital  configurations.
  Specifically, the central host of a polar-ring galaxy is the galaxy which has been
  inevitably transformed from a late-type spiral into an early-type S0 galaxy  during 
  the  merging.
  The ring component, on the other hand, is the `galaxy' which has been dramatically transformed
  from a late-type spiral into a narrow  ring or annuli  owing to the violent
  gravitational interaction and gaseous dissipation during  the merging.
  Although both 
  specific  orbital  configurations and gaseous dissipation in galaxy merging are required
  for the formation of polar-ring galaxies in the present  model,
  these constraints also  give natural explanations to
  observed trends  such as the prevalence of S0 among polar-ring galaxies (e.g., 
  Whitmore 1990),
  the rarity of the polar-ring galaxies  among S0 galaxies (e.g., Whitmore et al. 1991), 
  and an appreciably
  larger amount of interstellar gas in polar-rings (e.g., Sackett 1991).
  Moreover it  is found that
  the morphology of  polar-rings such as a narrow ring (e.g., ESO 415 - G 26), 
  annular rings (e.g., NGC 4650a),
  helical rings (e.g., NGC 2658), and double rings (e.g., ESO 474 - G 26) can reflect 
  both the orbital  parameters of galaxy merging and 
  the initial mass ratio of two merger precursor galaxies.
  Thus, the present study demonstrates that 
  a merger remnant of a gas-rich galaxy merger is 
  one of promising candidates of polar-ring S0 galaxies.

\acknowledgments
We are grateful to the referee Curtis Struck-Marcell for valuable comments, 
which greatly contribute to improve
the present paper.
K.B. thanks to the Japan Society for Promotion of Science (JSPS)
Research Fellowships for Young Scientist.

\newpage

\begin{deluxetable}{ccccccccc}
\footnotesize
\tablecaption{Model parameters \label{tbl-1}}
\tablehead{
\colhead{model no.} & \colhead{Orbit type} & 
\colhead{$m_{2}$} & \colhead{$r_{2}$} & \colhead{$M_{\rm g}$} &
\colhead{${\theta}_{1}$} & \colhead{$V_{\rm rel}$} & 
 \colhead{$r_{\rm p}$} & \colhead{Ring  morphology}}

\startdata
Model 1 & PO & 2.0 & 1.41 & 0.2 & 90.0 & 0.5 & - & narrow ring \\ 
Model 2 & PO & 2.0 & 1.41 & 0.2 & 60.0 & 0.5 & - & narrow ring \\ 
Model 3 & PO & 2.0 & 1.41 & 0.2 & 30.0 & 0.5 & - & no ring \\ 
Model 4 & PA & 2.0 & 1.41 & 0.2 & 90.0 & - & 0.2 & no ring \\ 
Model 5 & PO & 2.0 & 1.41 & 0.0 & 90.0 & 0.5 & - & no ring \\ 
Model 6 & PO & 2.0 & 1.41 & 0.2 & 90.0 & 1.5 & - & disturbed narrow ring \\ 
Model 7 & PO & 1.5 & 1.22 & 0.2 & 90.0 & 0.5 & - & narrow ring \\ 
Model 8 & PO & 1.0 & 1.00 & 0.2 & 90.0 &  0.5 & - & narrow ring \\ 
Model 9 & PO & 0.7 & 0.84 & 0.2 & 90.0 & 0.5 & -  & narrow ring \\ 
Model 10 & PO & 0.7 & 0.84 & 0.2 & 90.0 & 1.5 & -  & peculiar narrow ring \\ 
Model 11 & PO & 0.7 & 0.84 & 0.2 & 90.0 & 2.5 & -  & peculiar double rings \\ 
Model 12 & PO & 0.7 & 0.84 & 0.2 & 90.0 & 5.0 & -  & transient Cartwheel-like ring \\ 
Model 13 & PO  & 0.5 & 0.71 & 0.2 & 90.0 & 0.5 & - &  double rings \\ 
Model 14 & PO  & 0.5 & 0.71 & 0.2 & 60.0 & 0.5 & - &  double rings \\ 
Model 15 & PA  & 0.5 & 0.71 & 0.2 & 80.0 & - & 0.02 &  double rings \\ 
Model 16 & PO  & 0.3  & 0.55  & 0.2 & 90.0 & 0.5 & - & disk ($transient$ annular rings)  \\ 
Model 17 & PO  & 0.1 & 0.32 & 0.2 & 90.0 & 0.5 & - & disk (no ring) \\ 

\enddata
\end{deluxetable}

\newpage
\newpage
\figcaption{
  Initial  mass distribution of two 
  progenitor disks in a galaxy merger (left panel)
and final mass distribution of the merger remnant at $T$ = 14.0
in our units (right panel)
projected onto the  $xz$ plane
  for Model 1. A galaxy intrudes from the polar axis of the other
  one is referred to as the `intruder' whereas the other galaxy as the `victim'. An arrow indicates
  the direction of initial relative velocity of these two galaxies.
  In order to show more clearly the polar-ring component in the merger remnant,
we here plot only the gaseous component of the victim and the stellar
one  of the intruder in the right panel.
  Here, the halo component is not plotted in these two  panels.
\label{fig-1}}

\figcaption{Time evolution of merger orbit projected onto $x$ axis 
for Model 1.
Time is in units of $t_{\rm dyn}$.
Note that owing to dynamical friction of galaxy merging,
oscillation of the merger orbit along  $x$ axis is damped rapidly 
within $\sim$ 10 $t_{\rm dyn}$.
\label{fig-2}}

\figcaption{Time evolution of gas mass in our units for models with $m_{2}$
= 2.0 (Model 1)  and 0.3 (Model 16).
Time is in units of $t_{\rm dyn}$.
Note that the interstellar gas of merging galaxies is more rapidly consumed   
by star formation
in the model with larger  $m_{2}$ because of the stronger dynamical 
interaction of galaxy merging.
\label{fig-3}}

\figcaption{
 Morphological evolution of 
 the intruder (top) and the victim (stellar component: second from the top,
 gaseous one: third from the top, new stellar one: bottom)
 projected onto  the $xz$ plane
in Model 1. 
Time ($T$), indicated in upper left-hand corner of the top
frame, is in units of $t_{\rm dyn}$.
\label{fig-4}}

\figcaption{
The same as Figure 4 but $T$ = 8.0, 10.0, and 12.0. 
\label{fig-5}}

\figcaption{
 Morphological evolution of 
 the intruder (top) and the victim (stellar component: second from the top,
 gaseous one: third from the top, new stellar one: bottom)
 projected onto  the $yz$ plane
in Model 1. 
Time ($T$), indicated in upper left-hand corner of the top
frame, is in units of $t_{\rm dyn}$.
\label{fig-6}}

\figcaption{
The same as Figure 6 but for $T$ = 8.0, 10.0, and 12.0.
\label{fig-7}}

\figcaption{
The upper panel shows the
radial mass distribution of the intruder (solid line with open triangles)
and the victim (dotted  line with open squares) projected onto $xy$ plane
at $T$ = 14.0 in  Model 1.
In this upper panel, the initial exponential mass distribution of
the victim projected onto $yz$ plane (solid line) and that of the intruder 
projected onto $xy$ plane (dotted line) are also indicated. Note that
both the mass distribution of the intruder and that of 
the victim show deviation from the initial exponential mass distribution.
The lower panel shows the projected velocity profile 
along $x$ axis for the intruder (solid line) and the victim (dotted  one)
at $T$ = 14.0 in  Model 1.
Note that the intruder has  an appreciably larger  amount of global
rotation even after galaxy merging.
\label{fig-8}}

\figcaption{
Mass distribution of new stellar component formed before  $T$ = 8.0 
(upper panel) and after   $T$ = 8.0 (lower one) projected onto
$xz$ plane in Model 1.
Note that the mass distribution
of new stellar component formed before  $T$ = 8.0 looks more like spherical
whereas that of new stellar component after   $T$ = 8.0 looks more like disky.
\label{fig-9}}

\figcaption{
Time evolution of radial mass distribution of gaseous and new stellar components
in the victim projected onto $yz$ plane.
Each circle represents the relative magnitude of
the total mass within  $R_{i} <$ $r$ $< R_{i+1}$ ($i$ = 1 $\sim$ 20), 
where $r$ is the radius
($r^{2}=y^{2}+z^{2}$) from the center of the victim, and the $R_{i}$ is given
as 0.5 $\times$ ($i$-1) $\times$ $R_{\rm d}$.
The larger circle at each point in a given time indicates the larger mass
accumulated within each bin.
From the left to right, each set of circles represents the radial mass
distribution at  $T$ = 0.0, 2.0, 4.0, 6.0, 8.0, 10.0, and 12.0.
Note that owing to strong dynamical effects of the intruder on the victim
disk, a `hollow' in the mass distribution (the smaller
circle)  at $r$ = 0.45 in $T$ = 2.0 and at $r$ = 0.10 in $T$ = 4.0,
in our units, can be observed. 
These hollows imply that the radially propagating density wave is excited
by the intruder during galaxy merging.
Note also that these hollows are not observed at $T$ = 12.0.
This result implies that the radial density wave are damped by gaseous
dissipation at the later evolution of galaxy merging.
The larger circle at $r$ = 0.0 in $T$ $>$ 8.0 is the result of efficient mass-transfer
to the central region in dissipative galaxy merging (the so-called
`gas fueling').
\label{fig-10}}

\figcaption{
A collection of final global morphology of merger remnants (at 
$T$ =14.0) in  which 
the polar-ring component has not been reproduced successfully
(Model 3, 4, and 5).
In the left (for Model 3) and middle panels (for Model 4),
the upper three panels show the mass distribution of the stellar component
of the intruder and that of the gaseous component of the victim,
respectively.
In the right (for Model 5, the model with no gas),
the upper three panels show the mass distribution of the stellar component
of the intruder and that of the stellar  component of the victim,
respectively.
\label{fig-11}}

\figcaption{
A collection of final global morphology of merger remnants in models with
different mass ratio of merger precursor galaxies ($m_{2}$), $m_{2}$ =
1.5 (left), 1.0 (middle), and 0.7 (right)
for the stellar component of the intruder (upper),
the stellar component of the victim (second from the top),
the gaseous component of the victim (third from the top),
and the new stellar component of the victim (bottom).
\label{fig-12}}

\figcaption{
The same as Figure 12 but for the models with 
$m_{2}$ =
0.5 (left), 0.3 (middle), and 0.1 (right).
\label{fig-13}}

\figcaption{Long-term morphological evolution projected onto $xy$ plane
(upper three panels) and onto 
$xz$ one (lower three panels) in the merger remnant of Model 13. 
Time ($T$), indicated in the upper left-hand corner of the upper panels,
is in units of $t_{\rm dyn}$.
\label{fig-14}}

\figcaption{
 Morphological evolution of 
 the intruder (top) and the victim (stellar component: second from the top,
 gaseous one: third from the top, new stellar one: bottom)
 projected onto  the $xz$ plane
in Model 16. 
Time ($T$), indicated in upper left-hand corner of the top
frame, is in units of $t_{\rm dyn}$.
\label{fig-15}}

\figcaption{
The same as Figure 15 but $T$ = 8.0, 10.0, and 12.0. 
\label{fig-16}}

\figcaption{
 Morphological evolution of 
 the intruder (top) and the victim (stellar component: second from the top,
 gaseous one: third from the top, new stellar one: bottom)
 projected onto  the $yz$ plane
in Model 16. 
Time ($T$), indicated in upper left-hand corner of the top
frame, is in units of $t_{\rm dyn}$.
\label{fig-17}}

\figcaption{
The same as Figure 17 but for $T$ = 8.0, 10.0, and 12.0.
\label{fig-18}}

\figcaption{
The upper panel shows the
radial mass distribution of the intruder (solid line with open triangles)
and the victim (dotted  line with open squares) projected onto $xy$ plane
at $T$ = 14.0 in  Model 16.
In this upper pannel, the initial exponential mass distribution of
the victim projected onto $yz$ plane (solid line) and that of the intruder 
projected onto $xy$ plane (dotted line) are also indicated. Note that
the mass distribution of  
the victim shows clear deviation from the initial exponential mass distribution.
The lower panel shows the projected velocity profile 
along $x$ axis for the intruder (solid line) and the victim (dotted  one)
at $T$ = 14.0 in  Model 16.
Note that the intruder does not show remarkable  global rotation especially in
the inner part of the merger remnant 
after galaxy merging.
\label{fig-19}}



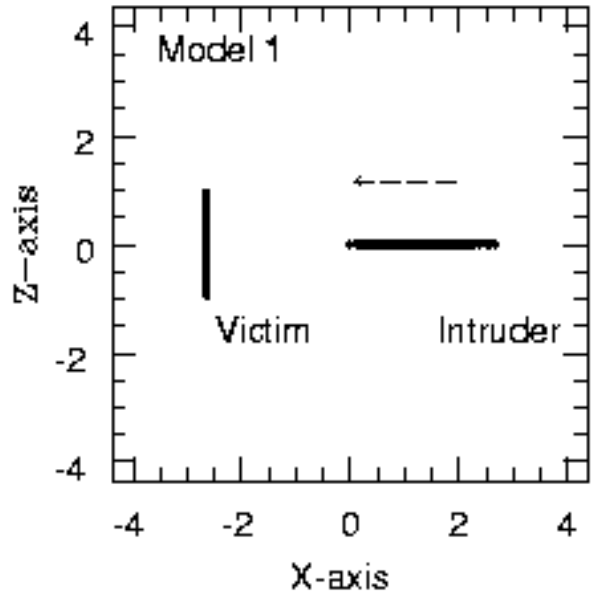
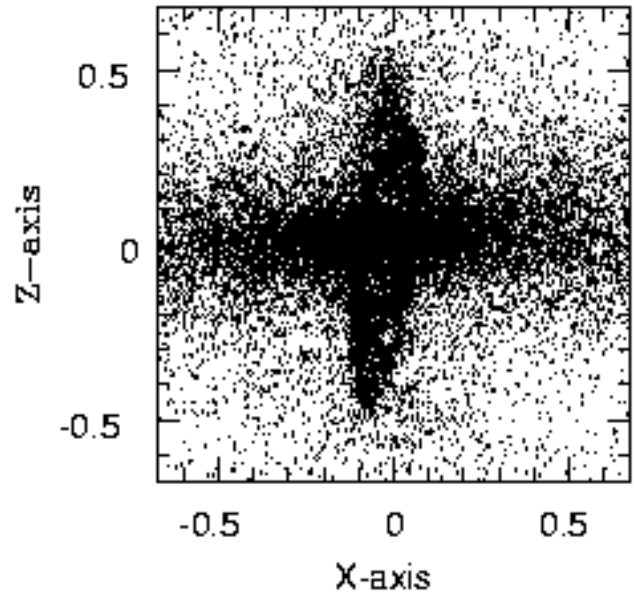
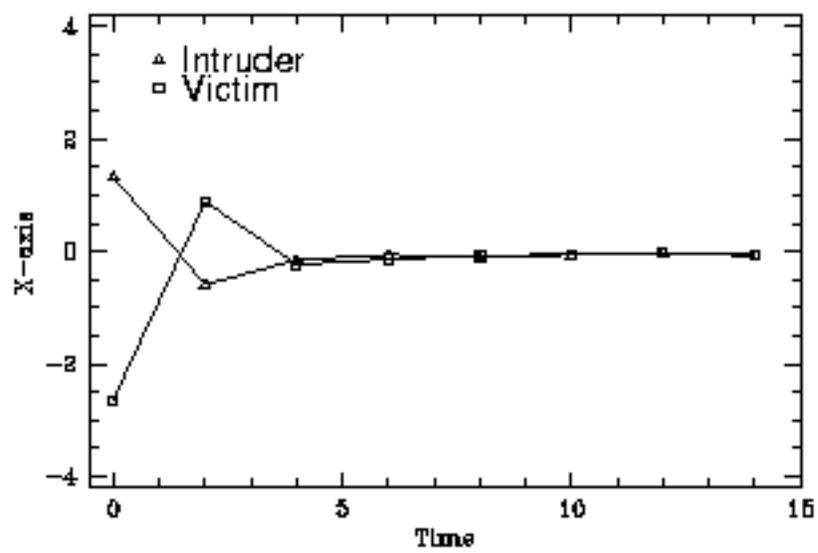
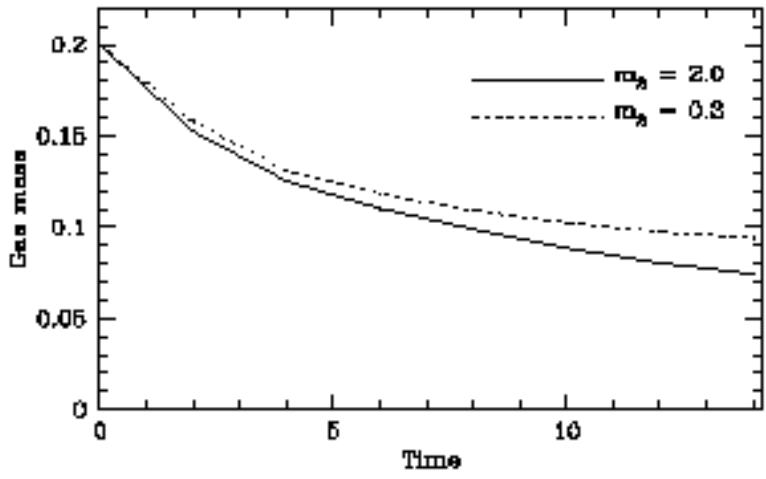
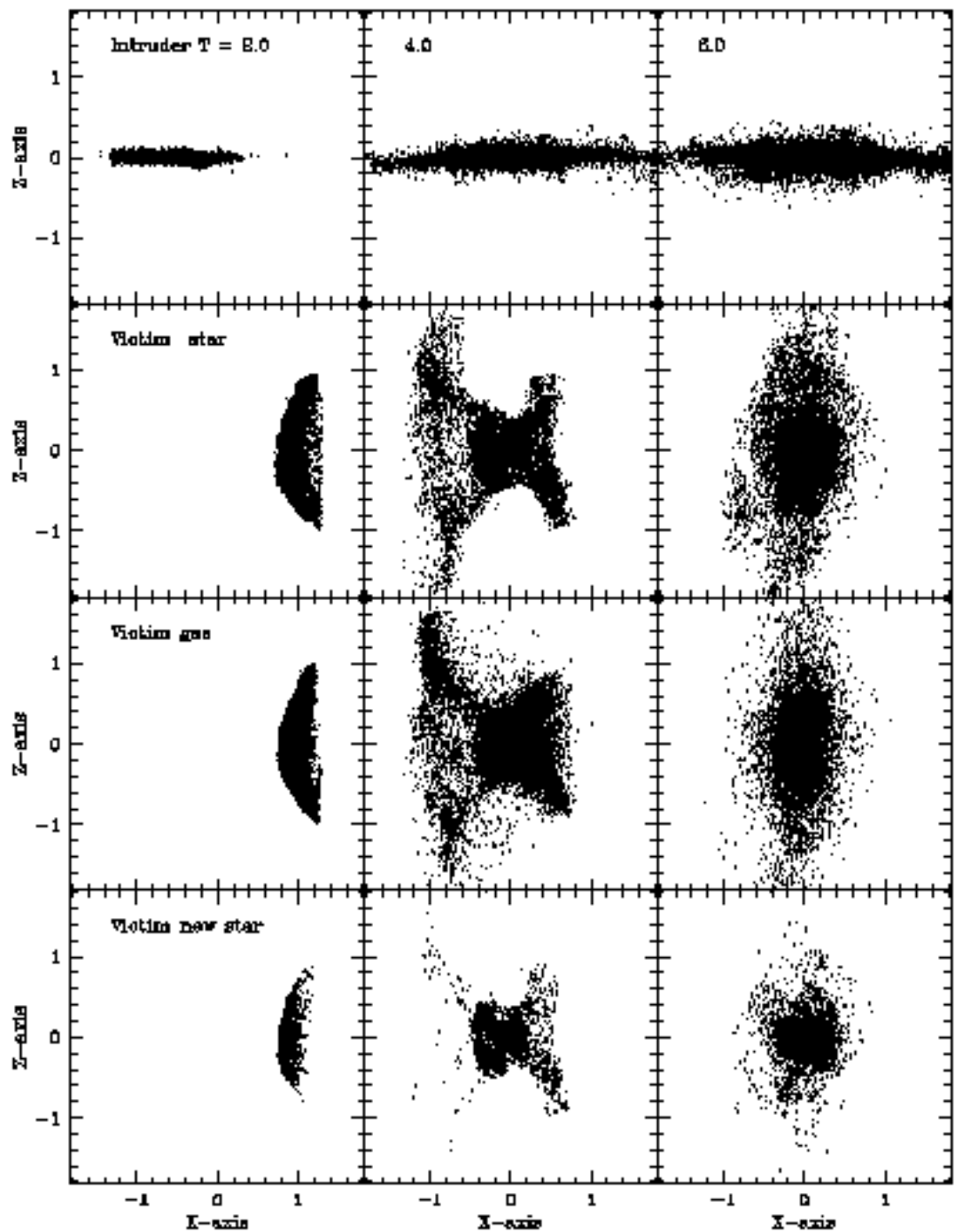
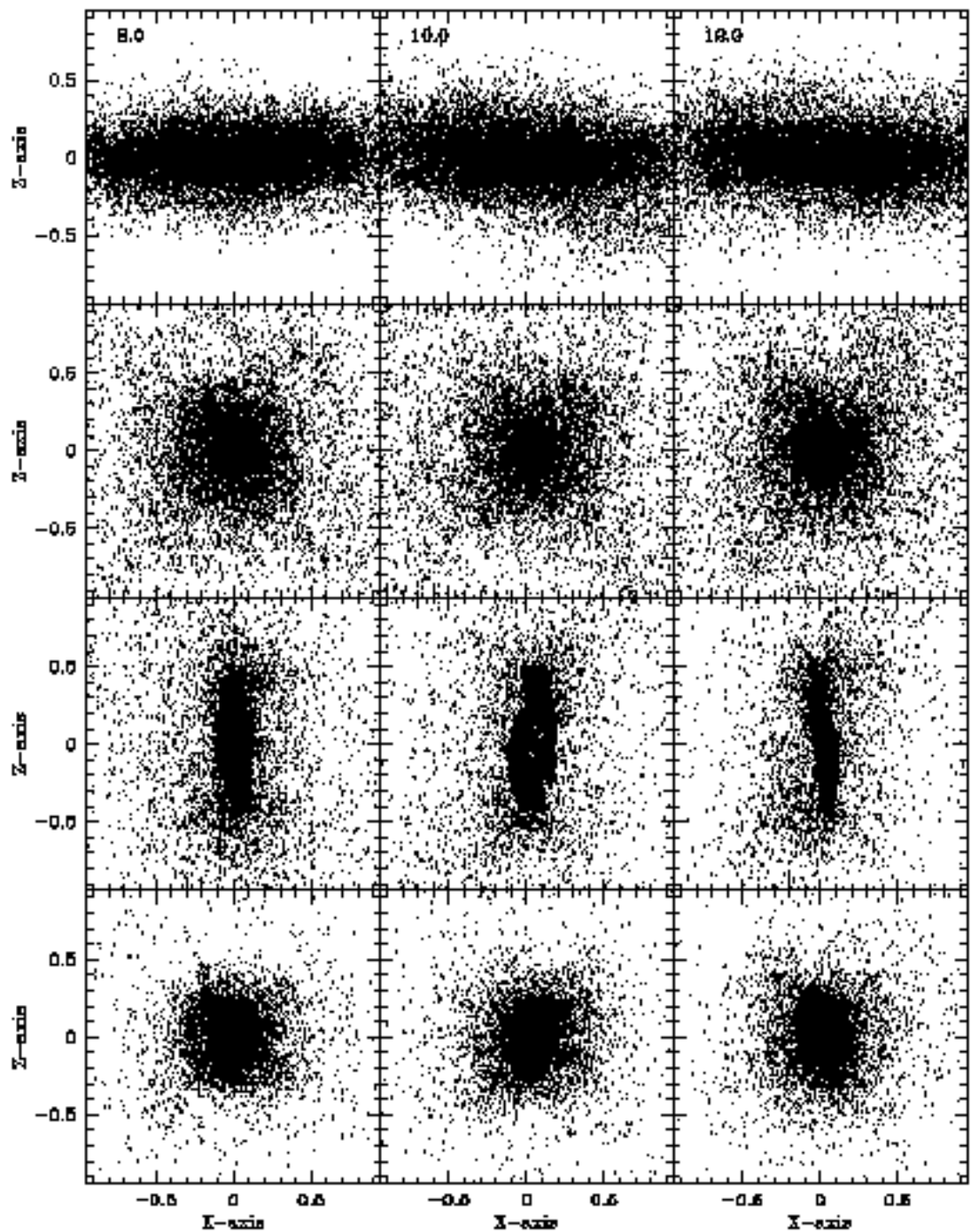
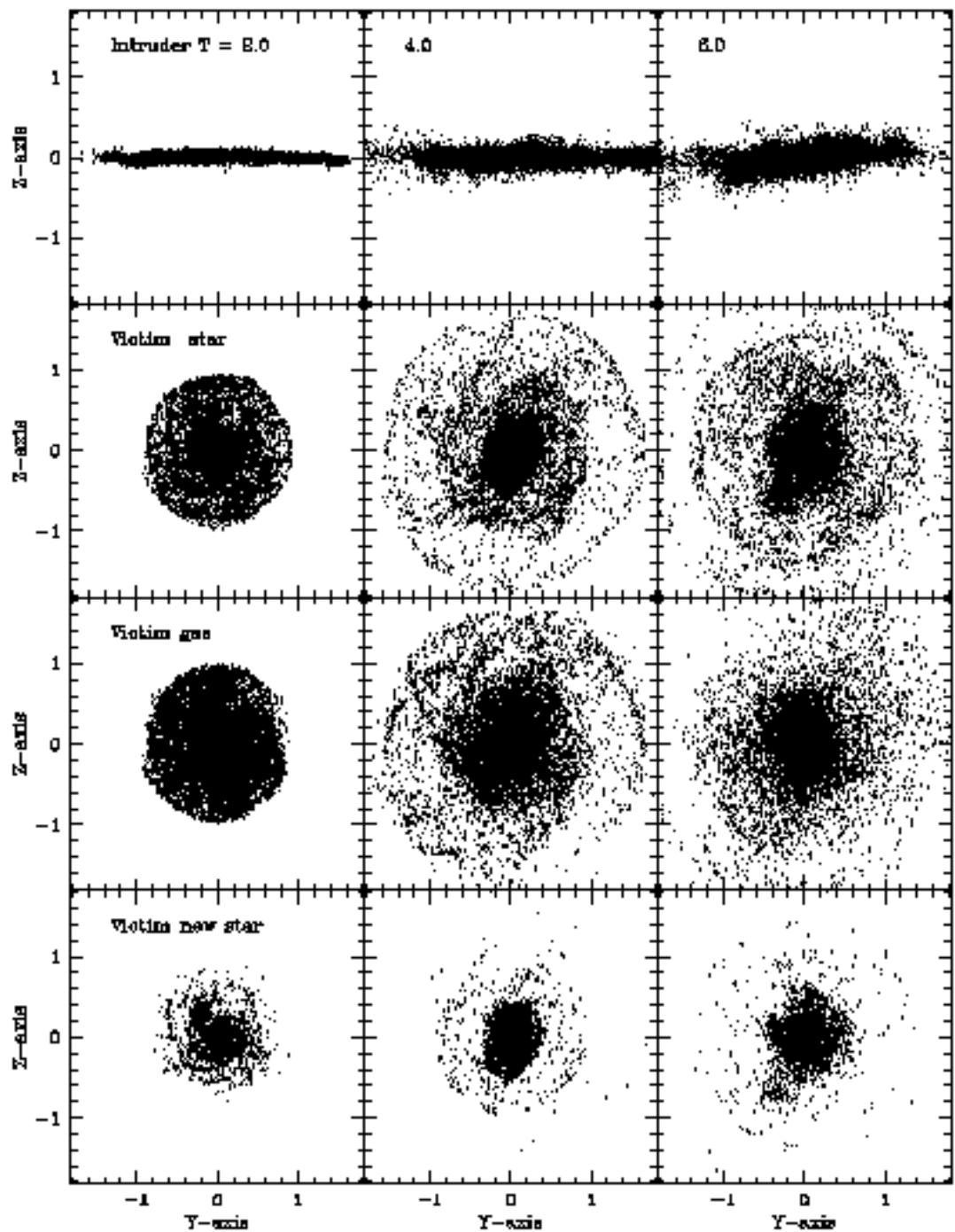
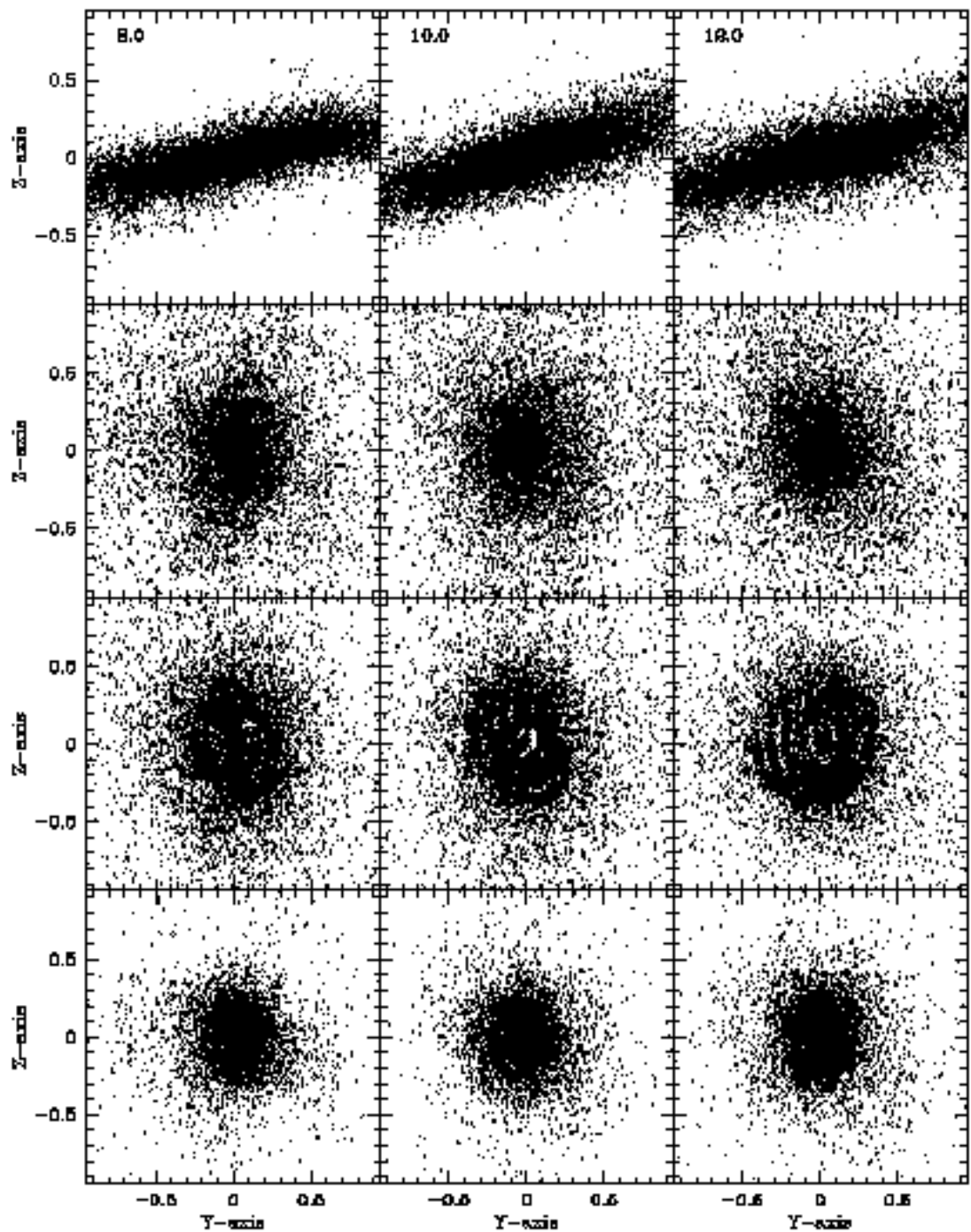
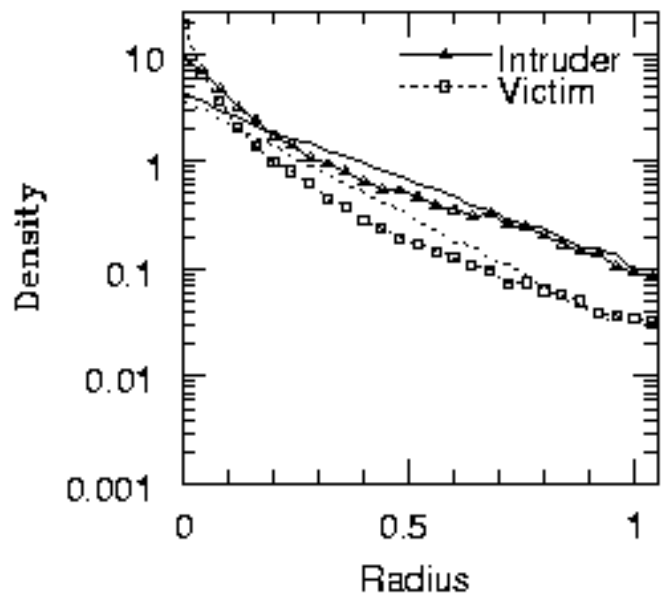
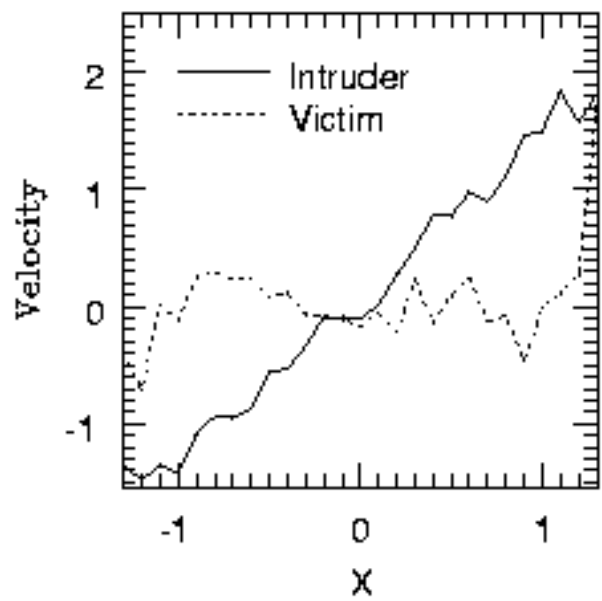
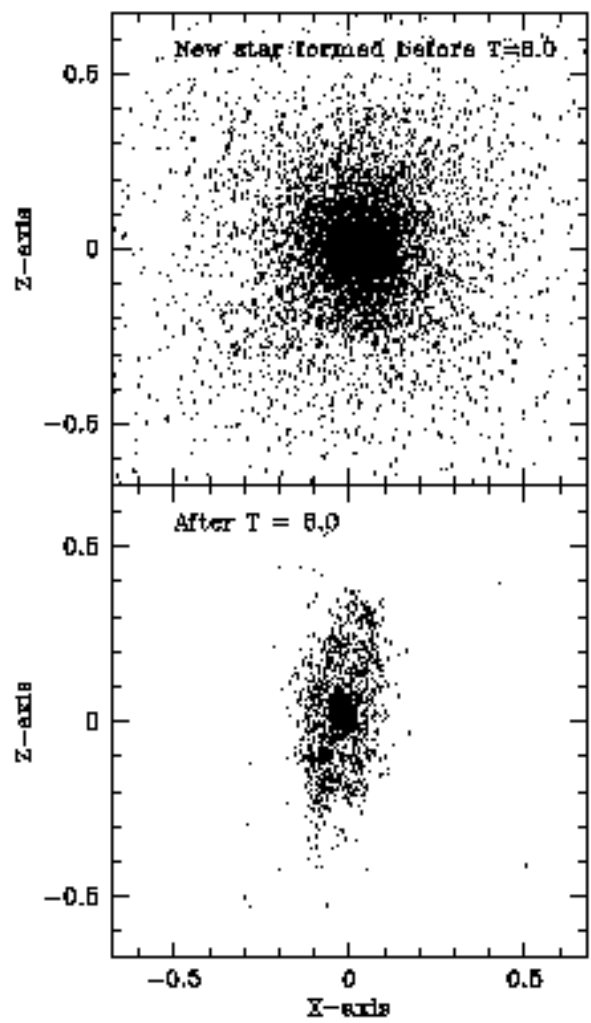
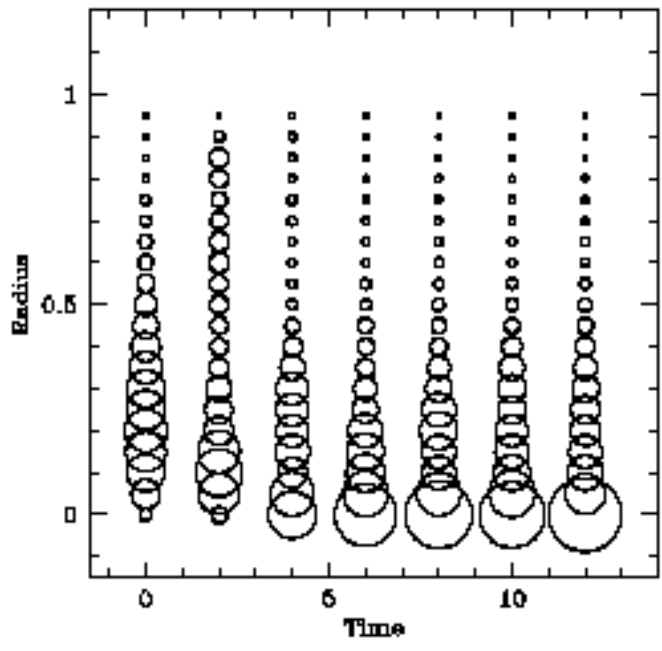
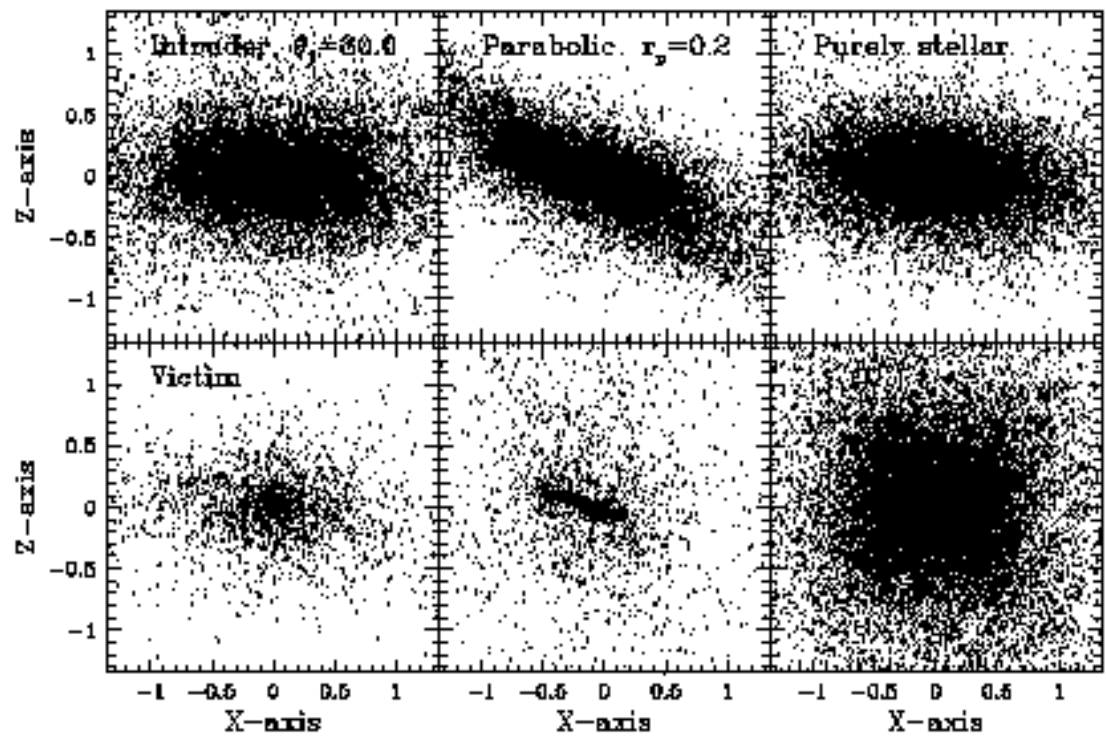
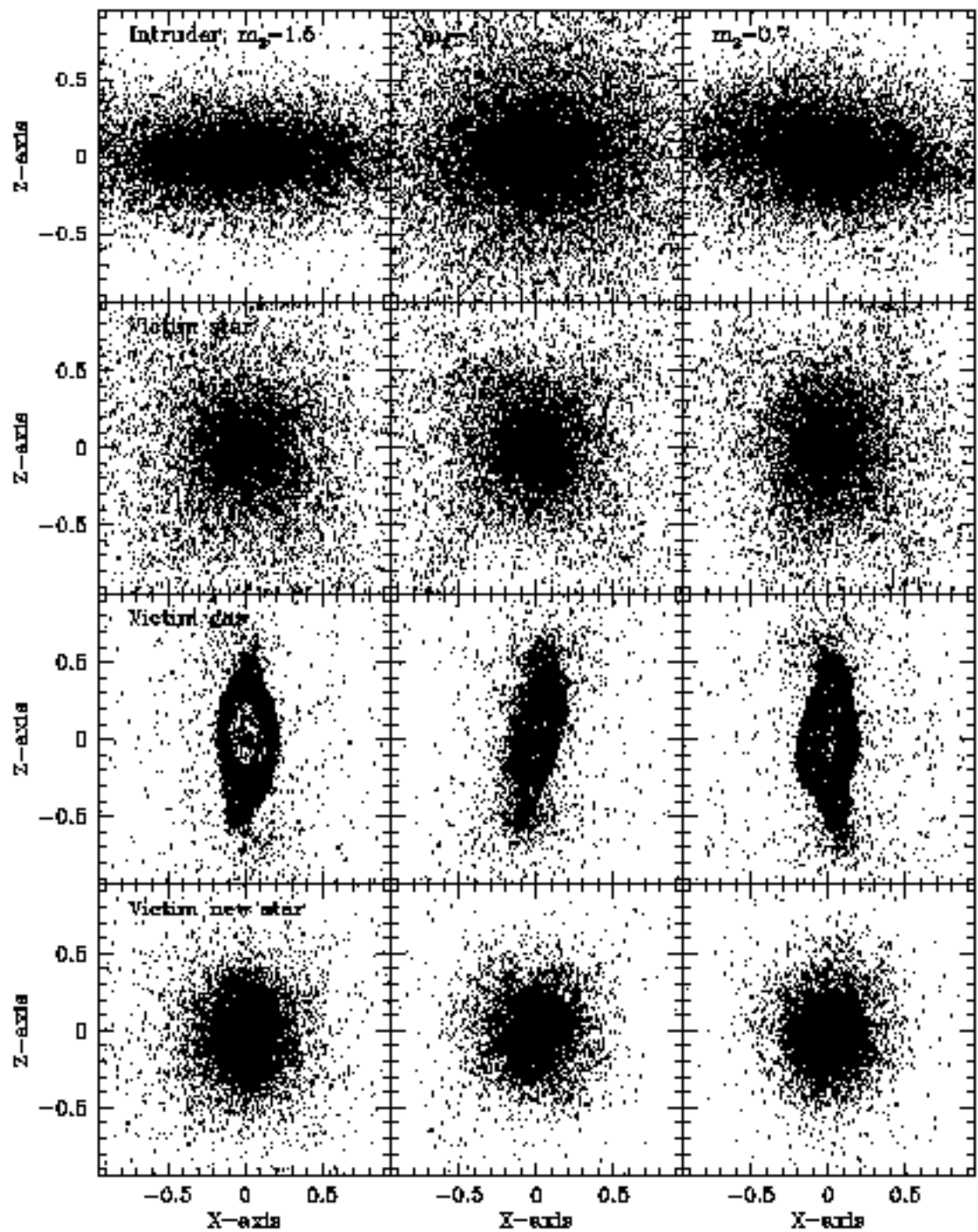
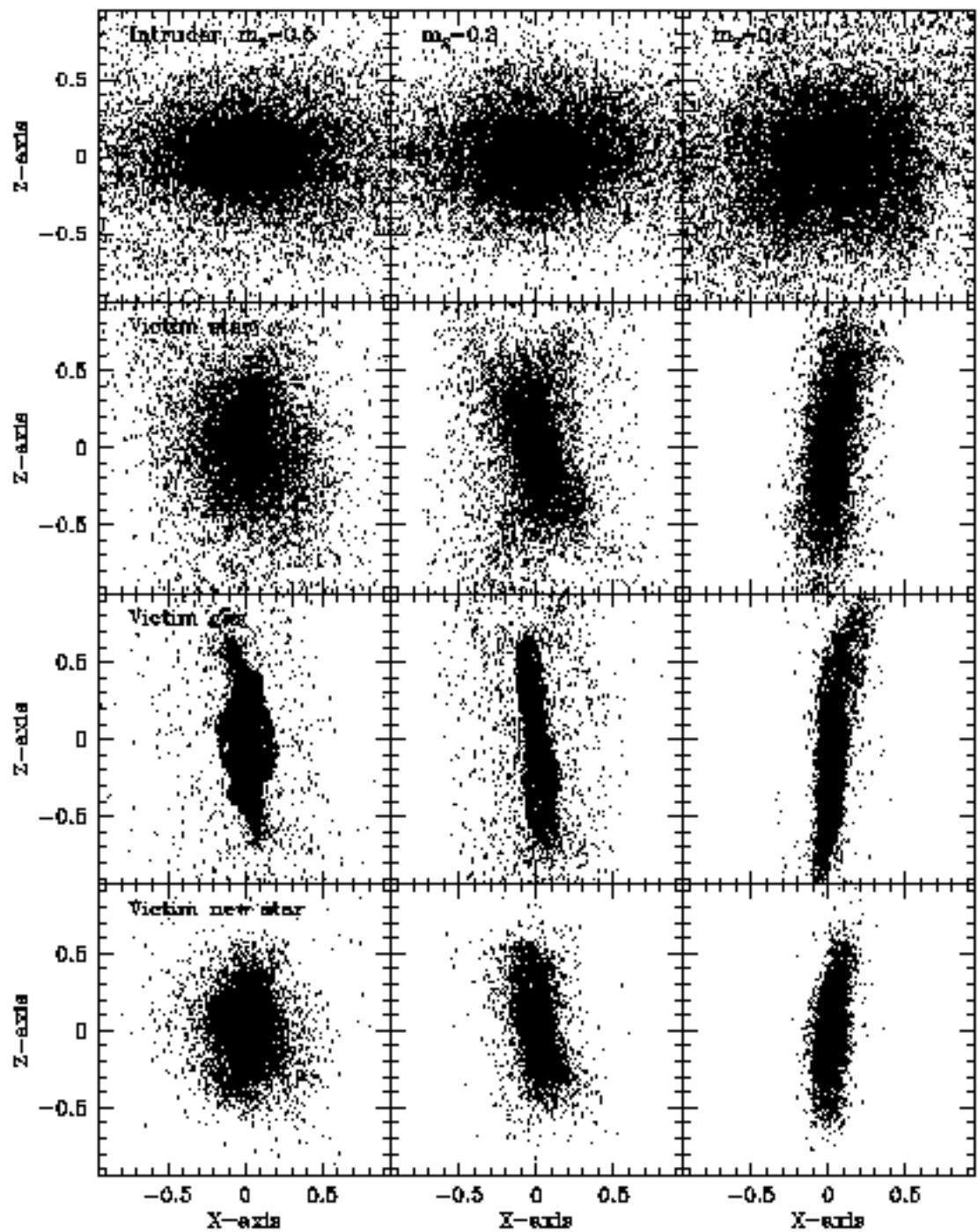
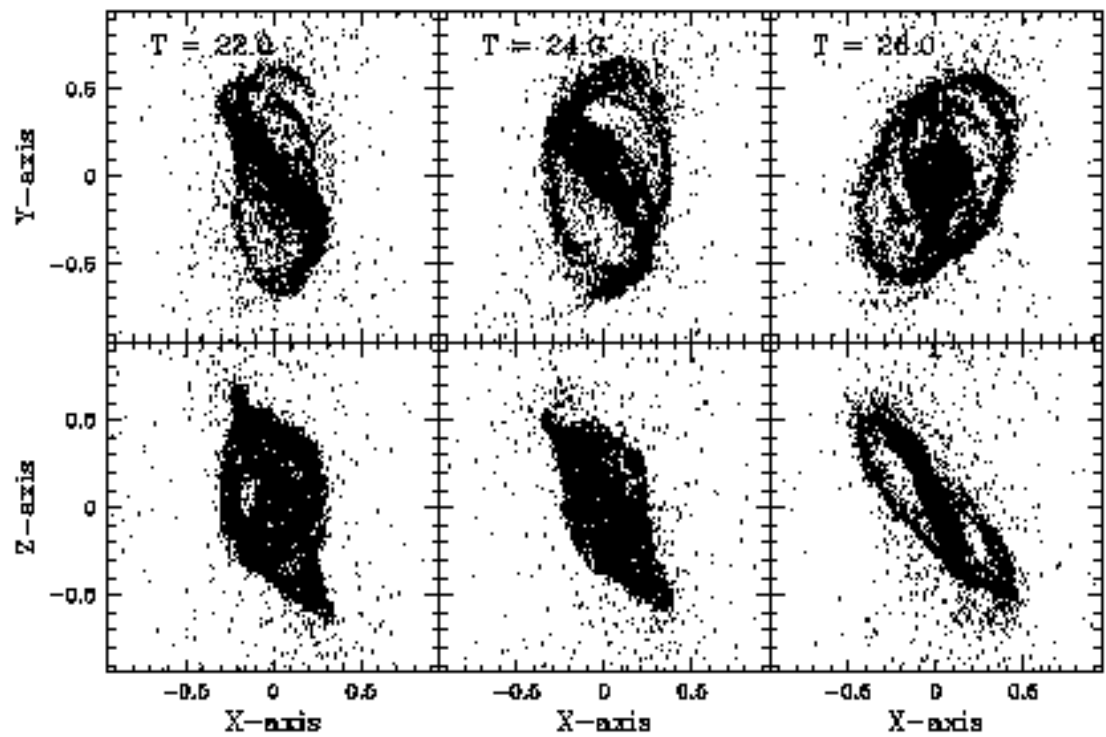
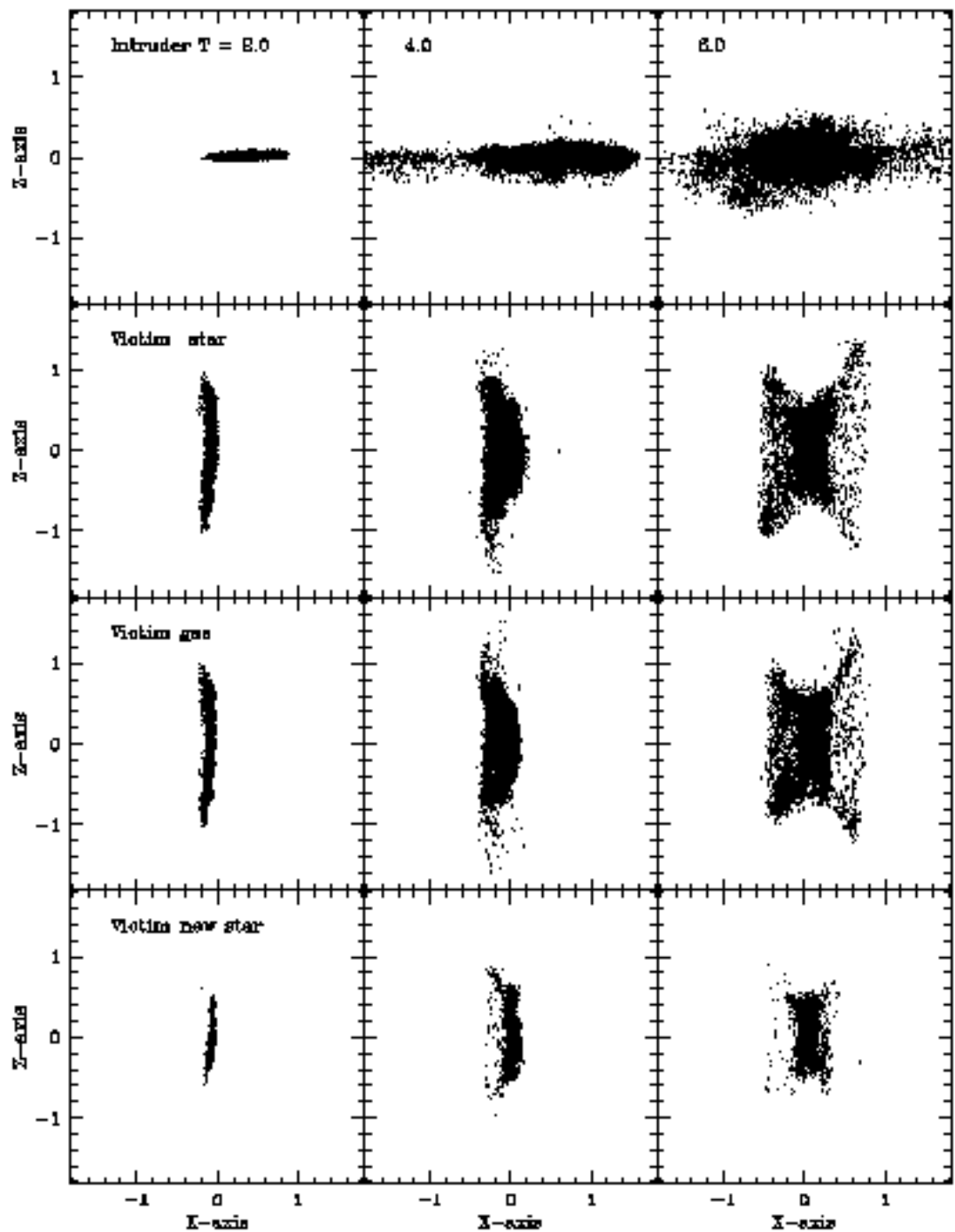
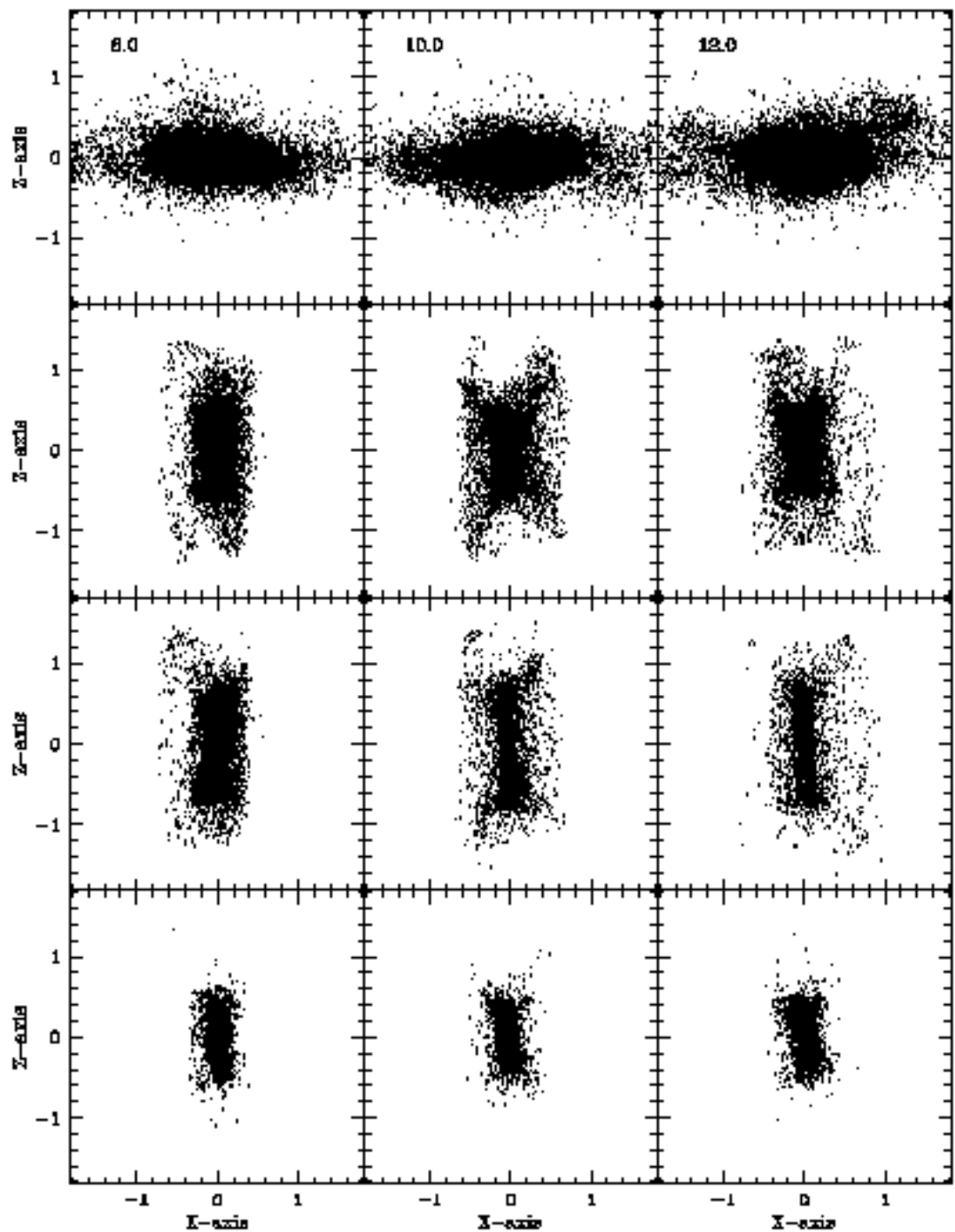
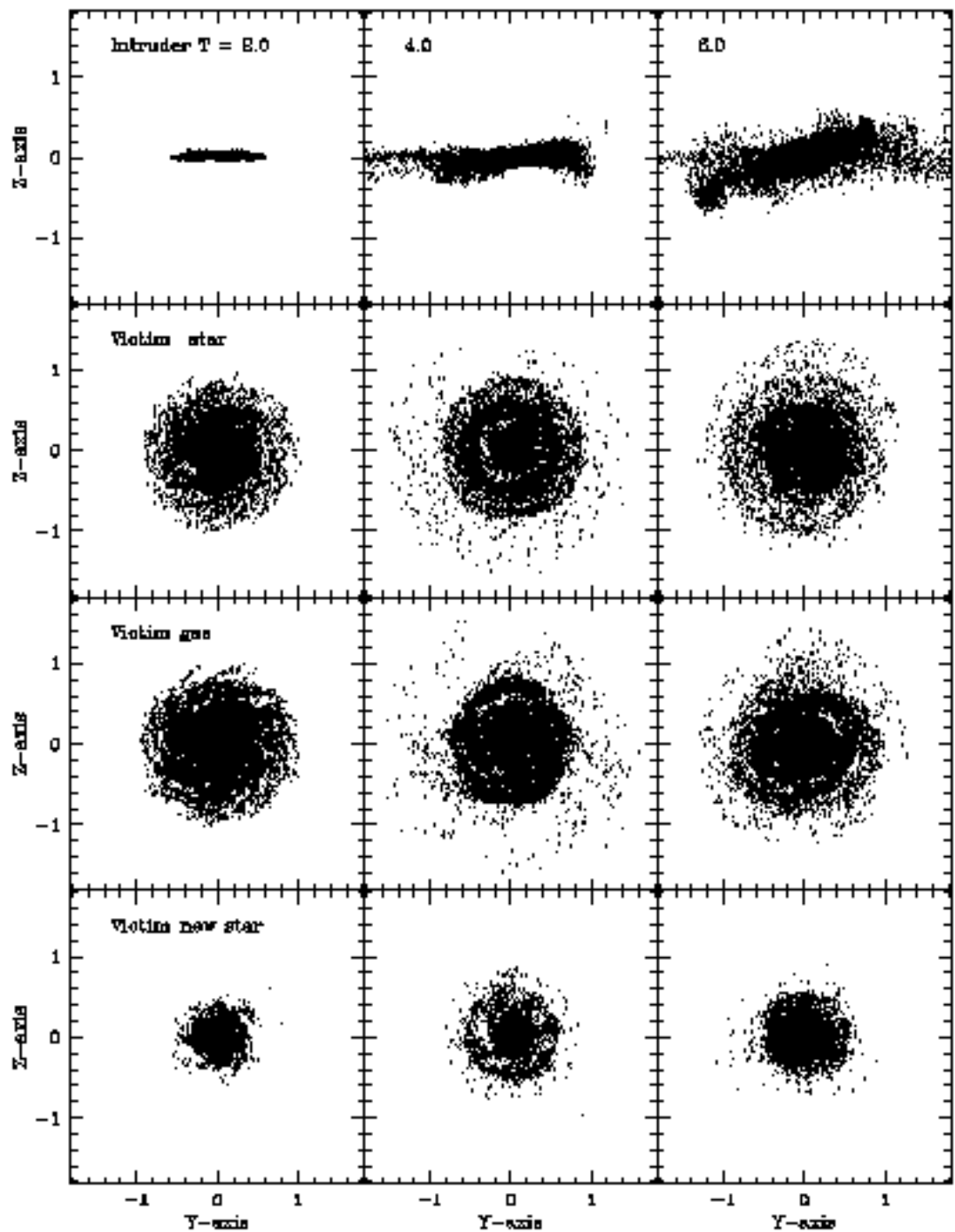
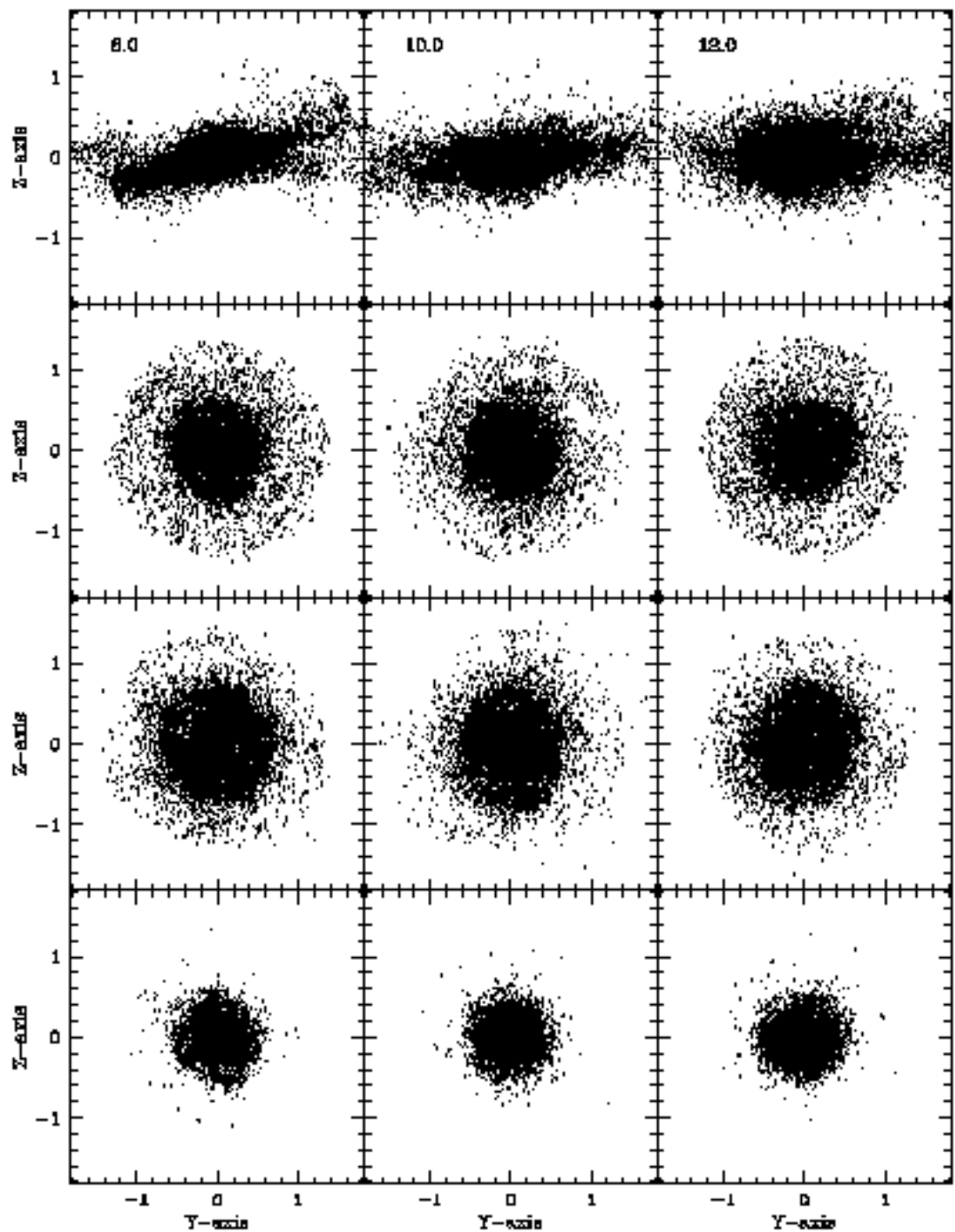
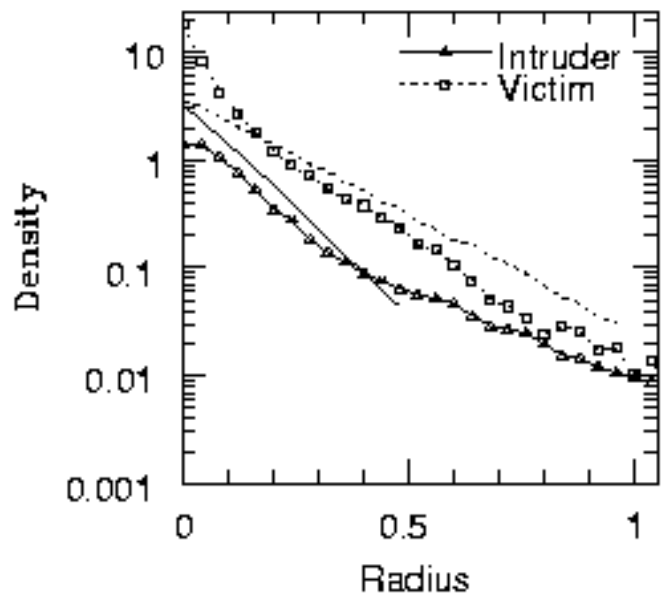
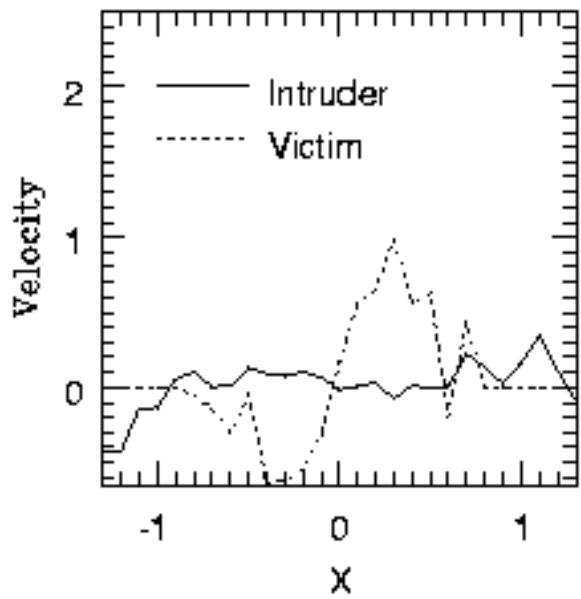
\end{document}